\documentclass[acmtog]{acmart}
\acmSubmissionID{345}

\usepackage{booktabs} 
\usepackage[title]{appendix}
\usepackage{wrapfig}
\usepackage[normalem]{ulem}

\theoremstyle{plain}
\newtheorem{thm}{Theorem}[section] 

\theoremstyle{definition}

\theoremstyle{proposition}
\newtheorem{prop}[thm]{Proposition}

\setcitestyle{square}

\usepackage[ruled]{algorithm2e} 

\SetAlFnt{\small}
\SetAlCapFnt{\small}
\SetAlCapNameFnt{\small}	
\SetAlCapHSkip{0pt}
\IncMargin{-\parindent}

\acmJournal{TOG}
\acmVolume{0}
\acmNumber{0}
\acmArticle{0}
\acmYear{0}
\acmMonth{0}

\begin{document}
\title{Surface2Volume: Surface Segmentation Conforming Assemblable Volumetric Partition} 

\author{Chrystiano Ara\'{u}jo $^*$}
\affiliation{\institution{University of British Columbia}}

\author{Daniela Cabiddu $^*$}
\affiliation{\institution{CNR IMATI}}

\author{Marco Attene}
\affiliation{\institution{CNR IMATI}}

\author{Marco Livesu}
\affiliation{\institution{CNR IMATI}}

\author{Nicholas Vining}
\affiliation{\institution{University of British Columbia}}

\author{Alla Sheffer}
\affiliation{\institution{University of British Columbia\\($^*$joint first authors)}}

%

\renewcommand\shortauthors{Ara\'{u}jo, C. et al}

\begin{teaserfigure}
   \centering
   \includegraphics[width=\linewidth]{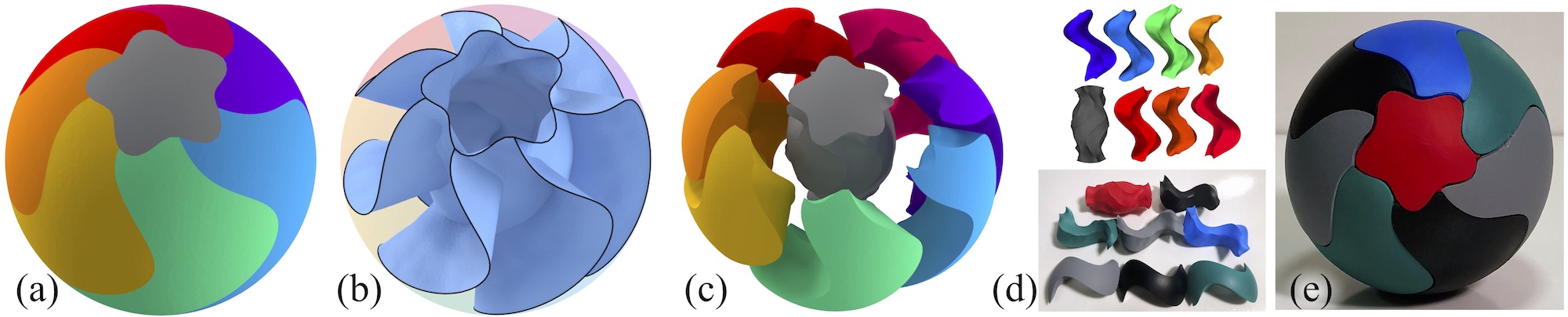}
   \caption{ Surface2Volume (left to right): \revised{(a)} Input multi-color surface; \revised{(b)} inner interfaces generated by Surface2Volume; \revised{(c)} assemblable parts conforming to the surface-segmentation; \revised{(d)} virtual \revised{(top)} and fabricated \revised{(bottom)} single-color parts; and \revised{(e)} assembled target object.}
   \label{fig:teaser}
\end{teaserfigure}

\begin{abstract}
Users frequently seek to fabricate objects whose outer surfaces consist of regions with different surface attributes, such as color or
material. Manufacturing such objects in a single piece is often challenging or even impossible. The alternative is to partition them into single-attribute volumetric parts that can be fabricated separately and then assembled to form the target object. Facilitating this approach requires partitioning the input model into parts that {\em conform} to the surface segmentation and that can be moved apart with no collisions. 
We propose {\em Surface2Volume}, a partition algorithm capable of producing such {\em assemblable} parts, each
of which is affiliated with a single attribute, the outer surface of whose assembly conforms to the input surface geometry and segmentation. 
In computing the partition we strictly enforce conformity with surface segmentation and assemblability, and optimize for ease of fabrication by minimizing part count, promoting part simplicity, and simplifying assembly sequencing.  
We note that computing the desired partition requires solving for three types of variables: \revised{per-part} assembly trajectories, partition topology, i.e. the connectivity of the interface surfaces separating the different parts, and the geometry, or location, of these interfaces.  
We efficiently produce the desired partitions by addressing one type of variables at a time: first computing the assembly trajectories, then determining interface topology, and finally computing interface locations that allow parts assemblability. 
\old{We iterate these steps for on complex models that necessitate assembly sequencing.}
\revised{We algorithmically identify inputs that necessitate sequential assembly, and partition these inputs gradually by computing and disassembling a subset of assemblable parts at a time.} 
We demonstrate our method's robustness and versatility by employing it to partition a range of models with complex surface segmentations into assemblable parts. We further validate our framework via output fabrication and comparisons to alternative partition techniques.
\end{abstract}

%
%
%

%
%




\newcommand{\cino}  [1]{{\color{magenta}	 Cino: #1}}
\newcommand{\jaiko}	[1]{{\color{orange}		 Marco: #1}}
\newcommand{\dany}	[1]{{\color{blue}		 Daniela: #1}}
\newcommand{\alla}	[1]{{\color{cyan}		 Alla: #1}}
\newcommand{\nick}	[1]{{\color{teal}		 Nick: #1}}
\newcommand{\chrys}	[1]{{\color{green} 		 Chrystiano: #1}}
\newcommand{\edit}	[1]{{\color{red}		 #1}}

\newcommand{\old} [1] {{}}
\newcommand{\revised} [1] {{\color{black} #1}}

\newcommand{\com} [1]{{}} 

\newcommand{\Obj}{\mathcal{O}} 

\maketitle



\section{Introduction}
\label{sec:intro}

\begin{figure}[t]
\includegraphics[width=\columnwidth]{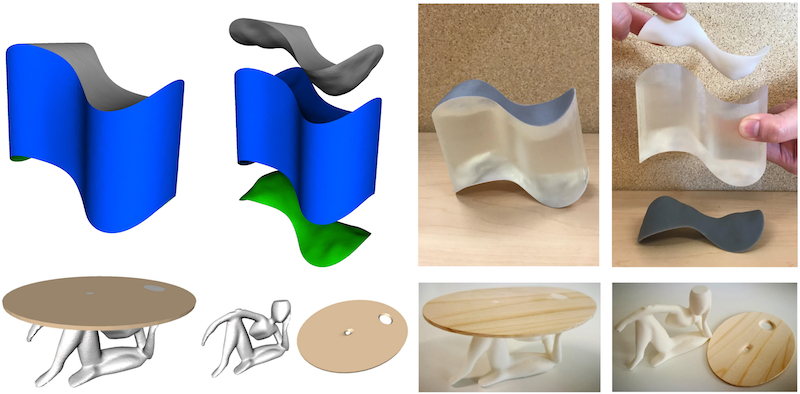}
\caption{Surface2Volume facilitates fabrication of multi-material objects whose parts are created using materials with different opacity (wavy cylinder, top), or ones necessitating the use of different fabrication technologies (table, bottom). Here we use FDM for the plastic base, and milling for the wooden top. The method produces a single part for the three disjoint regions labeled as plastic.}
\vspace{-15pt}
\label{fig:hybrid_fabrication}
\end{figure}

Digital fabrication algorithms \old{can} \revised{are} successfully \old{be} used to \old{fabricate} \revised{create} real-life replicas of virtual models with uniform color and material. However, users often wish to create objects with non-uniform visible surface attributes, such as shapes whose outer surface consists of regions with different color, opacity, or texture (Figures~\ref{fig:teaser},~\ref{fig:hybrid_fabrication}).  
Manufacturing such objects as a single solid necessitates the use of multi-attribute, or multi-material, fabrication methodologies, or after-the-fact surface painting.
These approaches exhibit numerous limitations (Section~\ref{sec:related}).
An appealing alternative is to decompose models with multi-attribute surfaces into single-attribute volumetric parts corresponding to the different surface regions \revised{(Figure~\ref{fig:teaser}c)}; fabricate these parts independently using widely available single-attribute fabrication tools \revised{(Figure~\ref{fig:teaser}d)}; and then assemble these parts together to form the desired object (Figure~\ref{fig:teaser}e). 
The algorithmic challenge in employing this multi-part fabrication approach is to obtain suitable model partitions, ones that {\em conform} to the input segmentation and allow post-fabrication part assembly. 
 We propose {\em Surface2Volume}, a new algorithm for computing such segmentation-conforming, assemblable partitions (Figure~\ref{fig:teaser}c). 

The partitions we compute associate each part with a single attribute and satisfy a number of key requirements. First, the computed parts strictly {\em conform} to the surface segmentation: the exposed, or outer, surface of the part assembly reproduces the 
input attribute-based segmentation. Second, the produced parts are {\em assemblable}: a user can take the separately manufactured parts and assemble them into an exact replica of the input model. Together, these two requirements ensure that the produced parts can 
indeed be used for multi-part fabrication. 
To simplify the assembly process, we also require the produced parts to allow for simple, {\em linear} assembly trajectories. We aim to keep the original surface segmentation intact when possible, avoiding unseemly seams through single-attribute regions, 
and seek to produce parts that allow for multiple assembly orders. Finally, to facilitate easy fabrication and assembly, we seek to produce parts with simple geometry, and avoid creating tiny, and thus hard to manipulate, parts. 

\begin{figure}
\includegraphics[width=\linewidth]{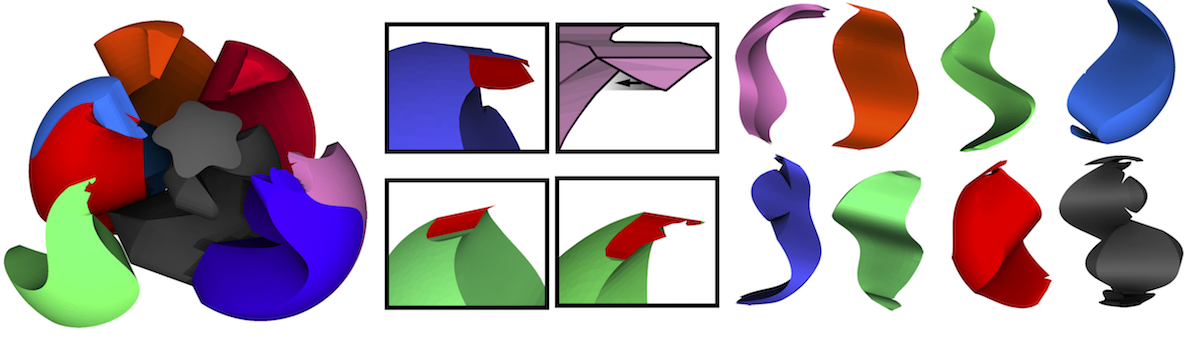}
\caption{The CSG engine used by Yao et al \protect \shortcite{yao2017interactive} produces an invalid, non-assemblable partition on an input which necessitates complex interface geometry (invalid parts highlighted in the close ups). Our result in Figure~\ref{fig:teaser}. }
\label{fig:swirl_ball}
\label{fig:yao1}
\end{figure}

Generating an assemblable partition given a fixed surface segmentation requires computing the connectivity and the geometry of the part {\em interfaces}, or interior surface boundaries between parts. The requirements above often necessitate forming interfaces with complex topology and non-trivial geometry (e.g. Figure~\ref{fig:teaser}b). Most prior methods for segmentation conforming assemblable partition are designed for restricted sets of segmentations and only consider a limited set of interface geometries and typologies; they consequently fail on more general inputs (Section~\ref{sec:related}). Yao et al.~\shortcite{yao2017interactive} generate assemblable partitions for interlocking furniture, targeting inputs whose interfaces are dominated by extrusions of surface region boundaries. 
This method fails to produce assemblable  parts when used on more general free-form models  (Section~\ref{sec:related}, Figure~\ref{fig:swirl_ball}).
{\em Surface2Volume} robustly partitions both free-form and man-made geometries, and is particularly well suited for inputs that require less regular and more free-form interfaces. 
 

Computing a desirable partition requires solving for three types of unknowns: assembly trajectories, interface topology and interface location.
We enable efficient and robust assemblable partition generation by developing an algorithm that efficiently computes 
these output properties sequentially rather than in tandem. 
We first predict per-part assembly trajectories that are likely to allow for assemblable partition by analyzing the input surface segmentation. 
We then generate an approximate partition that defines the topology and approximates the geometry of the part interfaces using a tetrahedral mesh of the input model as an underlying discretization. We formulate partitioning as a labeling problem, where labels correspond to parts and the partition energy reflects the properties 
we need to
satisfy (Section~\ref{sec:prob}). We find the desired partition using a classical graph-cut framework (Section~\ref{sec:labeling}). 
Finally, we optimize the geometry of the resulting interfaces strictly enforcing part assemblability (Section~\ref{sec:interface_smoothing}) using a specialized global-local solver.  

Our basic framework aims to partition objects into parts that can be assembled in any order
(e.g. Figure~\ref{fig:teaser})
and produces  a single part per surface region. On many inputs, the interaction between the initial surface regions, allows for only a subset of the corresponding parts to be disassembled right away (Figure~\ref{fig:update}). \old{We extend our method to handle these and other challenging} \revised{We handle such} inputs using a multi-pass partition process
\old{(Section~\ref{sec:overview})} \revised{(Section~\ref{sec:input_update})}. We support segmentation refinement (Figure~\ref{fig:dragon}, Section~\ref{sec:split}) when necessary to allow a valid partition, and  enable grouping of disjoint regions sharing the same attribute values into common parts (Figure ~\ref{fig:hybrid_fabrication}\revised{, bottom}, Section~\ref{sec:unify_components}), reducing part count and simplifying fabrication. 

We validate Surface2Volume by applying it to a wide range of inputs, manufacturing an array of multi-attribute objects using \old{our}  \revised{the resulting} partitioned geometries as input, and comparing our results  to prior art (Section~\ref{sec:results}). 
Our comparisons demonstrate that Surface2Volume robustly \old{handle}\revised{partitions} a vast array of  inputs including those that fail prior methods.
\revised{It produces the minimal possible part count on all the partitioned inputs and performs region splitting only when no alternative solution, independent of method, exists - namely when input surface regions associated with different attributes interlock (Figure~\ref{fig:dragon}).}

Our overall contribution is a robust and efficient algorithm that takes as input closed surface models  whose surfaces are segmented into regions associated with different attribute values and produces single-attribute assemblable parts that conform to the input segmentation.
Our method is particularly well suited for partition of natural shapes which can only be partitioned using  irregular, free-form interface surfaces.

\section{Background and Related Work}
\label{sec:related}

We build on existing research in a number of domains, \revised{reviewed below}. \old{including digital fabrication, shape segmentation, and assemblable partition.}

\paragraph{Multi-Material Fabrication.}
High-end 3D printers, such as Z Corp or HP Jet Fusion, allow simultaneous deposition of materials with different colors or mechanical properties; however they are extremely expensive and can only combine a limited set of materials. 
Emerging research into hybrid technologies that exploit off-the-shelf components to reduce costs~\cite{multifab} is not mature enough for a wide-scale adoption. 
Two-headed \revised{Fused Deposition Modeling (FDM)} machines can deposit two colored filaments at a time, but cannot support models where three or more colors or materials appear in the same slice; moreover, using this hardware may introduce color artifacts which persist even when specially optimized machine toolpaths for each head are used \cite{hergel14,reiner2014dual}. 
Most commodity fabrication hardware operates on one material at a time, using a single filament for FDM 3D printing or carving a single solid material block in a subtractive setting. Our method is designed to allow users to produce artifact-free objects consisting of any number of distinct materials using these widely accessible systems.

A recent line of research investigates methods for painting the surfaces of manufactured objects, including computational hydrographic printing \cite{ZhangHydrographic,DanieleHydrographics} and computational thermoforming \cite{schuller2016computational}. These works operate by simulating the behaviour of a printed sheet of ink or plastic under deformation, and are primarily suitable for near-convex, genus-0 objects. Outputs generated this way tend to fade over time and lose their coloring due to wear and tear. Our method has no convexity or genus constraints, and our outputs better retain their coloring as they consist of solid single material parts.




\paragraph{Shape Segmentation.} Most surface and volume segmentation methods focus on the computation of surface
charts or volumetric parts with specific intrinsic part properties and do not account for interactions between them~\cite{shamir2008survey,Sharp:2018:VSC,ho2012volume,strodthoff2017automatic}. Enforcing assemblability requires accounting for the interaction between parts, necessitating a more global solution methodology. 

\paragraph{Assemblable Partitions for Fabrication.} Volumetric partitioning has been extensively used to overcome a range of manufacturing hardware constraints and widen the range of fabrication techniques applicable to a given input~\cite{Cignoni:2016,LEMLA17}. 
Examples include breaking models into parts which are small enough to fit into a printer's chamber \cite{song2015printing,Luo:2012:CPM,hao2011efficient,vanek2014packmerger}; generating parts that can be efficiently packed~\cite{attene2015shapes,chen2015dapper,Yao:2015:LPP};  
\revised{construction of new models by an assemblage of parts from a database~\cite{Funkhouser} or via explicit construction of interlocking assemblies~\cite{DESIA}}; or enabling the use of specific fabrication technologies~\cite{MLSSP18,herholz2015approximating,metamolds,corecavity}.

Our work complements these approaches in its focus on partitioning for surface-attribute-driven multi-material fabrication.

The ability to assemble parts together to form the target model is a critical requirement for any technique which decomposes models for subsequent fabrication. Existing methods utilize a range of strategies to satisfy this requirement. 
Multiple methods use part assemblability as the key criterion for volumetric partition~\cite{song2012recursive,xin2011making,lo20093d,song2015printing,fu2015computational,Yao:2015:LPP}, and have the surface segmentation emerge from this partition. 
Other methods partition the volume and the surface simultaneously using strategies that ensure assemblability. For example, models partitioned using cut-through planes~\cite{Luo:2012:CPM,chen2015dapper,attene2015shapes,hildebrand2013orthogonal,hu2014approximate} can always be assembled using an assembly order that inverses the cutting sequence. 
Some methods such as ~\cite{song2015printing} can accommodate some constraints on the resulting surface segmentation, for instance avoiding segment boundaries in certain areas; however, none of the methods above is designed to incorporate prescribed surface segmentation boundaries.

\paragraph{Segmentation Conforming Assemblable Partitioning.}
Zhang et al. \shortcite{zhangassembling} partition 2D polygons into assemblable parts conforming to an outline segmentation. As they acknowledge, it is not clear how to extend their method to 3D space.
Several methods create shallow volumetric parts that correspond to surface segments which satisfy different restrictive sets of desirable properties, such as limited normal variation \cite{wang2016improved,MLSSP18,herholz2015approximating}, near planar, simple inter-segment boundaries \cite{song2016cofifab,wang2016improved}, or bounded size~\cite{song2016cofifab,vanek2014packmerger}.
Given the surface segmentations, they define the parts by extruding the segments either along a fixed direction to form flat-based prisms~\cite{wang2016improved,MLSSP18} or along the inverse normal direction to form shells \cite{herholz2015approximating,song2016cofifab,vanek2014packmerger}. As the authors acknowledge, these approaches can easily fail given more general surface segmentations. 
\setlength{\intextsep}{1pt} 
\setlength{\columnsep}{4pt} 
\begin{wrapfigure}{l}{0.25\linewidth}
  \begin{center}
   \includegraphics[width=\linewidth]{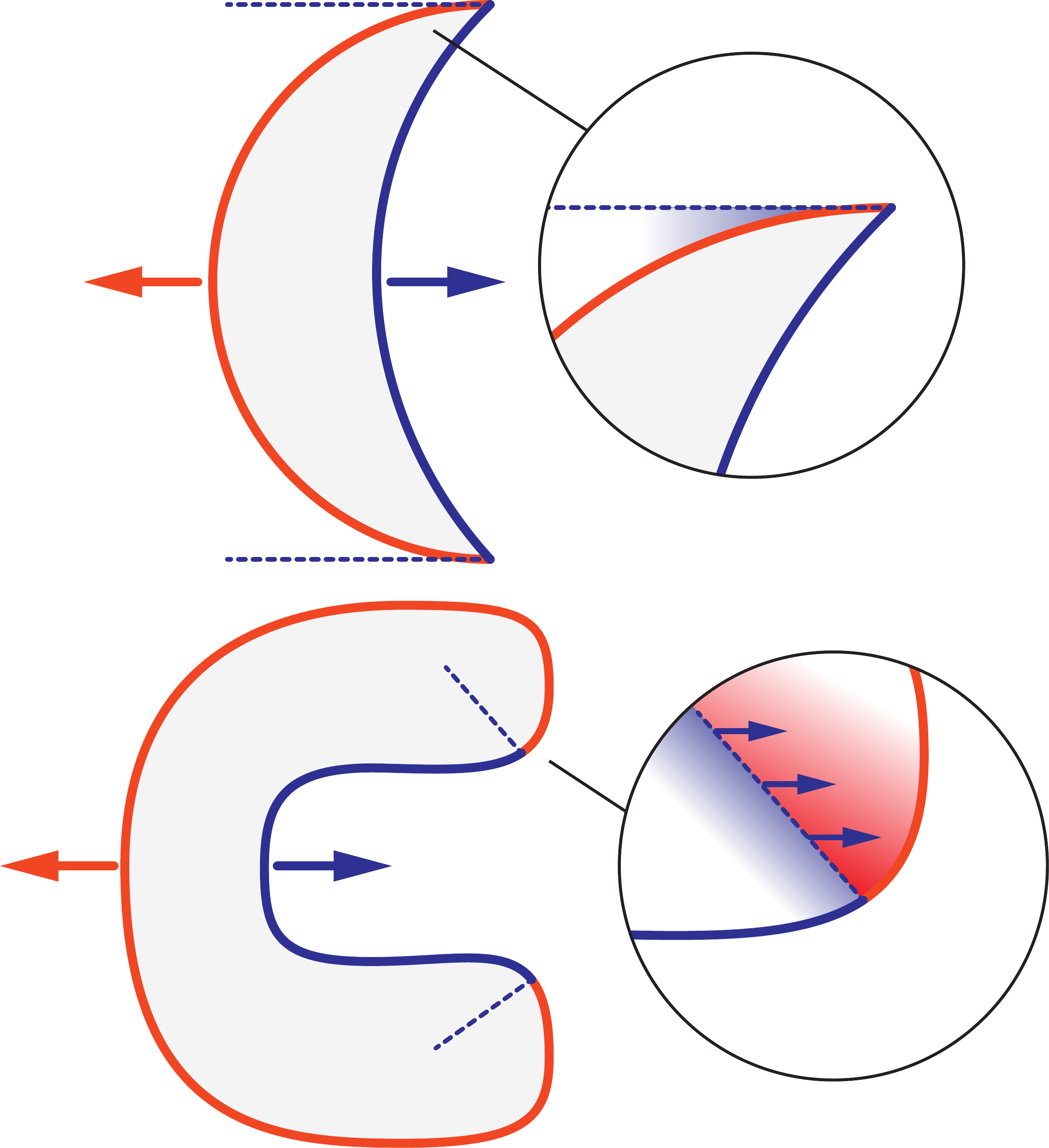}
  \end{center}
\end{wrapfigure}
Offsetting a part along a constant direction fails on even simple 2D inputs (see inset, top). Regardless of the offsetting direction, the blue part will span outside of the domain. Offsetting inwards along the surface normal direction does not work for concave objects -- parts interlock and do not allow assembly (inset, bottom).
We impose no restrictions on the input segmentation and successfully generate assemblable partitions that conform to input segments with large normal variation and complex non-planar boundaries (e.g. Figures~\ref{fig:swirl_ball},~\ref{fig:mosaic}). 


Yao et al. \shortcite{yao2017interactive} propose a partition method for interlocking furniture design. This method is best suited
for engineered shapes where the desired interfaces are dominated by linear extrusions. 
While it can extend to simple free-form geometries, it fails to partition inputs such as the soccer ball or milk-jug  (Figure~\ref{fig:mosaic}) and produces an invalid partition on the swirl ball, Figure~\ref{fig:swirl_ball} (our result is in Figure~\ref{fig:teaser}). Our framework complements this approach in its focus on more free-form input geometries, and robustly handles such inputs.

\paragraph{Assembly Path Computation.} A vast body of work, e.g.~\cite{joskowicz1991computational,joskowicz1999computer,wolter1991automatic,agrawala2003designing}, addresses computation of assembly paths or evaluates assemblability of given part configurations. While these methods analyze existing parts we focus on computation of part geometries that allow assembly.  

\revised{\paragraph{Reconstruction from Slices.} A final related area of work are planar-to-volume interpolation methods, which naturally occur in the case of reconstructing volumes from sliced data such as medical tomography. In these works (e.g.~\cite{LiuSurface,Bermano}), planar slices are marked with an attribute function and interpolated throughout the reconstructed volume. These methods trivially extend from planar-to-volume interpolations to surface-to-volume interpolations. While some of these works do attempt to make topological guarantees (e.g.~\cite{Lazar}), these guarantees are restricted to surface and genus and do not consider the problem of assemblability.}


\section{Problem Statement and Overview}
\label{sec:overview}

\subsection{Problem Statement}
\label{sec:prob}

\old{
The input to Surface2Volume is a 3D object $\mathcal{O}$ whose outer surface is segmented into several regions associated with different attributes. The goal of the algorithm is to partition this object into a set of volumetric parts that satisfy the following validity constraints and exhibit the desirable characteristics detailed below. 
}

\revised{The input to Surface2Volume is a closed 3D surface that is segmented into several regions associated with different attributes. The algorithm partitions the volumetric object $\mathcal{O}$ enclosed by this surface, into a set of volumetric parts that satisfy the following validity constraints and exhibit the desirable characteristics detailed below.}

\paragraph{Partition Validity.}
Validity requires the output parts to satisfy {\em segmentation conformity} and {\em linear assemblability} constraints. 
Conformity necessitates that each part $\Obj_i \in \mathcal{O}$ is associated with a single attribute, and that the visible portions of parts associated with each attribute exactly conform to the region(s) $\mathcal{R}_i$  corresponding to this attribute on the outer surface of the input object. Assemblability means that there exists a sequence of collision-free part trajectories that 
allows moving the parts away from one another until they are no longer in contact, or vice versa moving them from disjoint positions back into the assembly.
To simplify the assembly process and the computation, we restrict the allowable assembly trajectories to linear motions.
\revised{The linearity constraint also makes the output more tolerant to fabrication inaccuracies; assembly along non-linear paths is likely to be more sensitive to fabrication imprecisions.}
 In addition to these requirements, to be manufacturable the parts need to be manifold and self-intersection free. 

Formally, a collection of parts $\Obj_i \in \mathcal{O}$ is  {\em linearly assemblable} if there exists an order $\Obj_0,\ldots,\Obj_n$ \revised{and a direction $d_i$}  such that each part $\Obj_i$ can be {\em extracted}, or moved, along the direction $d_i$ from the sub-assembly $\Obj_i,\ldots,\Obj_n$ without intersecting any of the remaining parts. 
One of the key observations behind our method is that this brute-force assessment 
 of part extractability can be replaced by the assessment of the following two criteria, illustrated in the inset below: {\em interface extractability} and {\em region extractability}. 

\setlength{\intextsep}{1pt} 
\setlength{\columnsep}{4pt} 
\begin{wrapfigure}{r}{0.25\linewidth}
  \begin{center}
    \includegraphics[width=\linewidth]{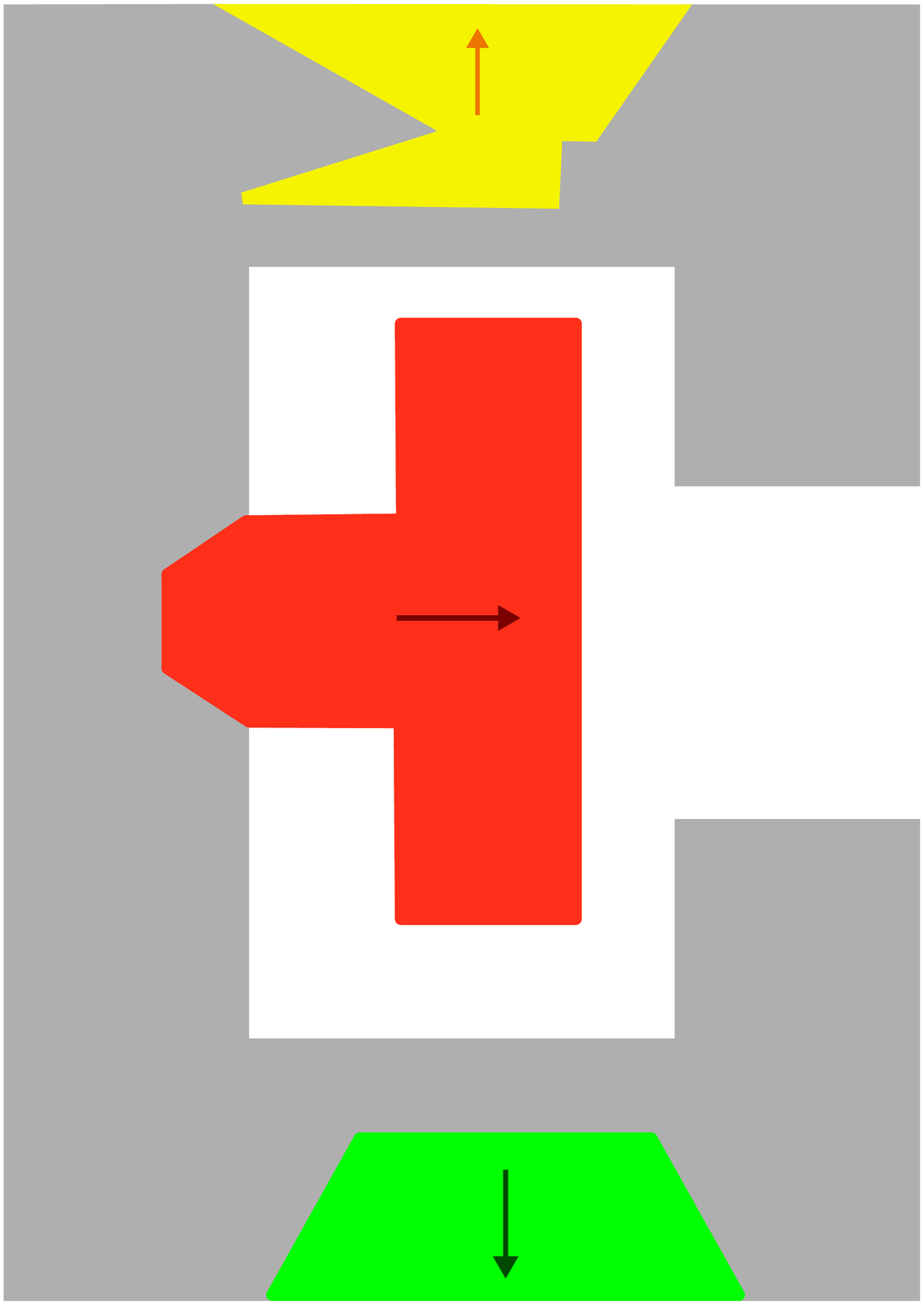}
  \end{center}
\end{wrapfigure}
We define the {\em interface} between two parts $\Obj_i$ and $\Obj_j$ as the union of points contained in both parts. 
Let $\mathcal{P}_i$ be the set of all the points on the outer surface of $\Obj_i$ which are {\bf not} on a shared interface with  $\Obj_j,~j > i$. We say $\Obj_i$ is \emph{\old{surface}\revised{region} extractable \revised{along a direction $d_i$}} if a ray shot from any point $p \in \mathcal{P}_i$ along $d_i$ does not intersect any other part $\Obj_j,~j > i$. 
We say $\Obj_i$ is \emph{interface extractable along the direction $d_i$} if the normal vector $n_p$ at any point $p$ on the interface between $\Obj_i$ and any of $\Obj_j,~j > i$ satisfies  $n_p \cdot d_i \leq 0 $ (i.e. if they form a non-acute angle). 
\revised{For brevity, we omit the direction $d_i$ where possible in the paper, and simply say that a part is \old{surface}\revised{region} extractable if some direction exists that it is \revised{region} \old{surface} extractable along; similarly for interface extractability.}
A part is extractable if it is both \old{surface}\revised{region} extractable and interface extractable (See Appendix \ref{app:proof} for a proof). In the inset, the yellow and green parts are \old{surface}\revised{region} extractable along the indicated directions; and red and green are interface extractable.  Consequently only the green part is fully extractable. 
Note that \revised{region} \old{surface} extractability can be assessed using the surface segmentation as input alone, and does not depend on the volumetric partition. \revised{This observation allows us to first compute extractable per-region directions and to then focus on forming volumetric parts that conform to these regions and satisfy interface extractability with respect to these directions.}



\begin{figure}
\includegraphics[width=\linewidth]{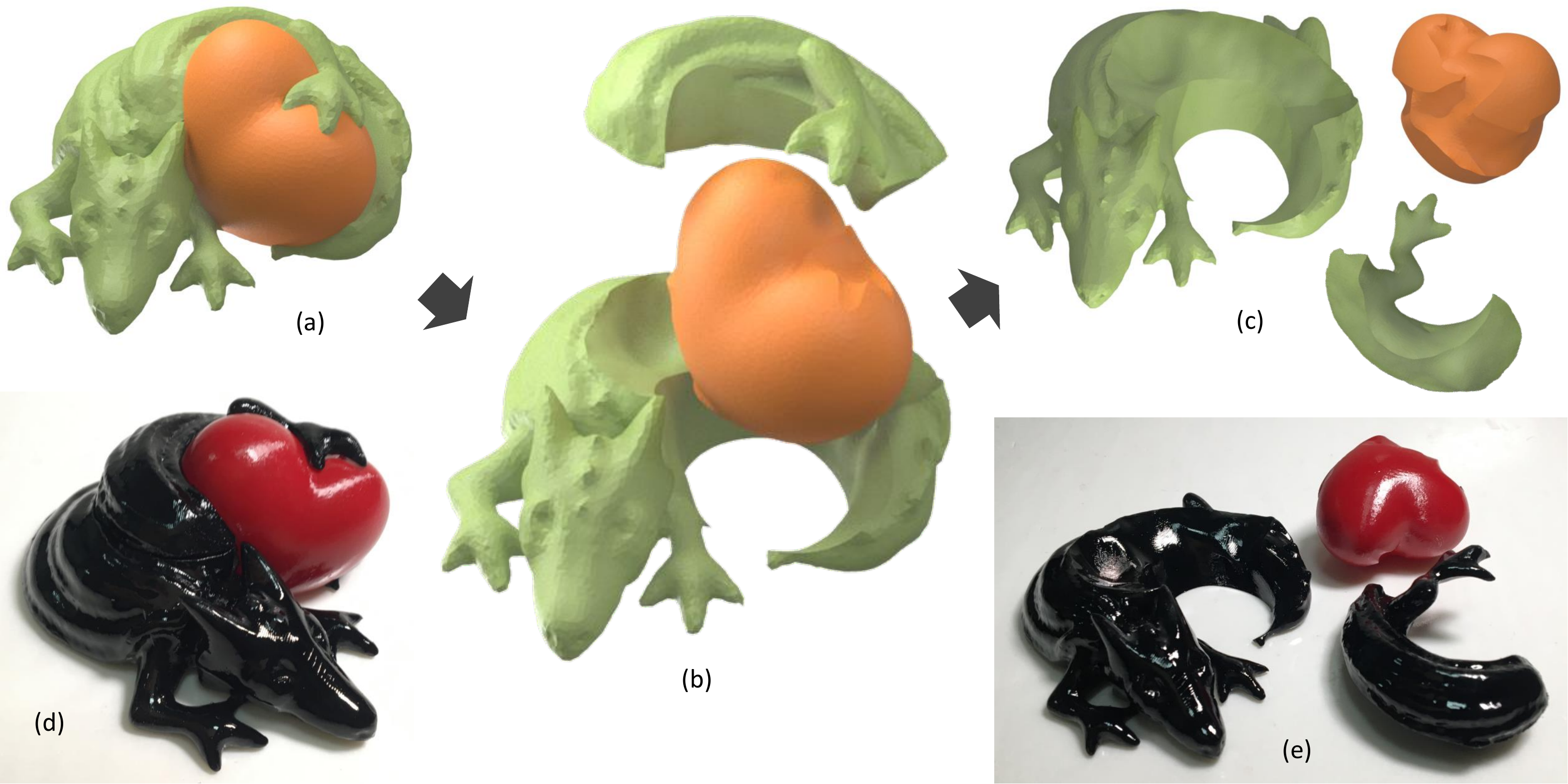}
\caption{Interlocking configurations: the surface regions on this input are not \old{surface} \revised{region} extractable, but splitting the dragon's body into two regions allows for a valid partition.
}
\label{fig:blocking1}
\label{fig:dragon}
\end{figure}

\paragraph{Partition Characteristics}
Among all valid partitions, we prefer ones which are easiest to manufacture and assemble. We aim to keep the number of parts as small as possible, ideally producing only as many parts as there are distinct attributes, and refine the surface segmentation only when no parts are otherwise extractable (Figure~\ref{fig:dragon}).
 We also seek to maximize the number of parts that can be extracted, or disassembled, simultaneously at each disassembly step. Both preferences are motivated by the desire to simplify the assembly process from a user perspective. \revised{
 In particular, simultaneous extractability makes the assembly process more self-evident, avoiding the need for complex assembly instructions. It also make parts more tolerant to fabrication inaccuracies. Single order (puzzle-like) assembly can fail due to inaccuracies in the critical key insertion stage. Both constraints } are explicitly accounted for in our partition algorithm design. Finally, to simplify both assembly and fabrication we search for parts with balanced sizes and smooth interfaces.  Parts with smoother interfaces are easier to manufacture and assemble, while very small parts are hard to manipulate. 

\paragraph{Discretization}
We define and subsequently optimize the desired partition properties using a tetrahedral mesh of the input model as an underlying discretization. 
For simplicity's sake, the formulation presented here and through Sections~\ref{sec:aeap_decomp}-\ref{sec:splitting} addresses the scenario where  each contiguous input surface region has a different attribute value.
We discuss the extension to scenarios where the number of attributes is smaller than the number of regions in Section~\ref{sec:unify_components}.

\setlength{\intextsep}{1pt} 
\setlength{\columnsep}{4pt} 
\begin{wrapfigure}{r}{0.25\linewidth}
  \begin{center}
    \includegraphics[width=\linewidth]{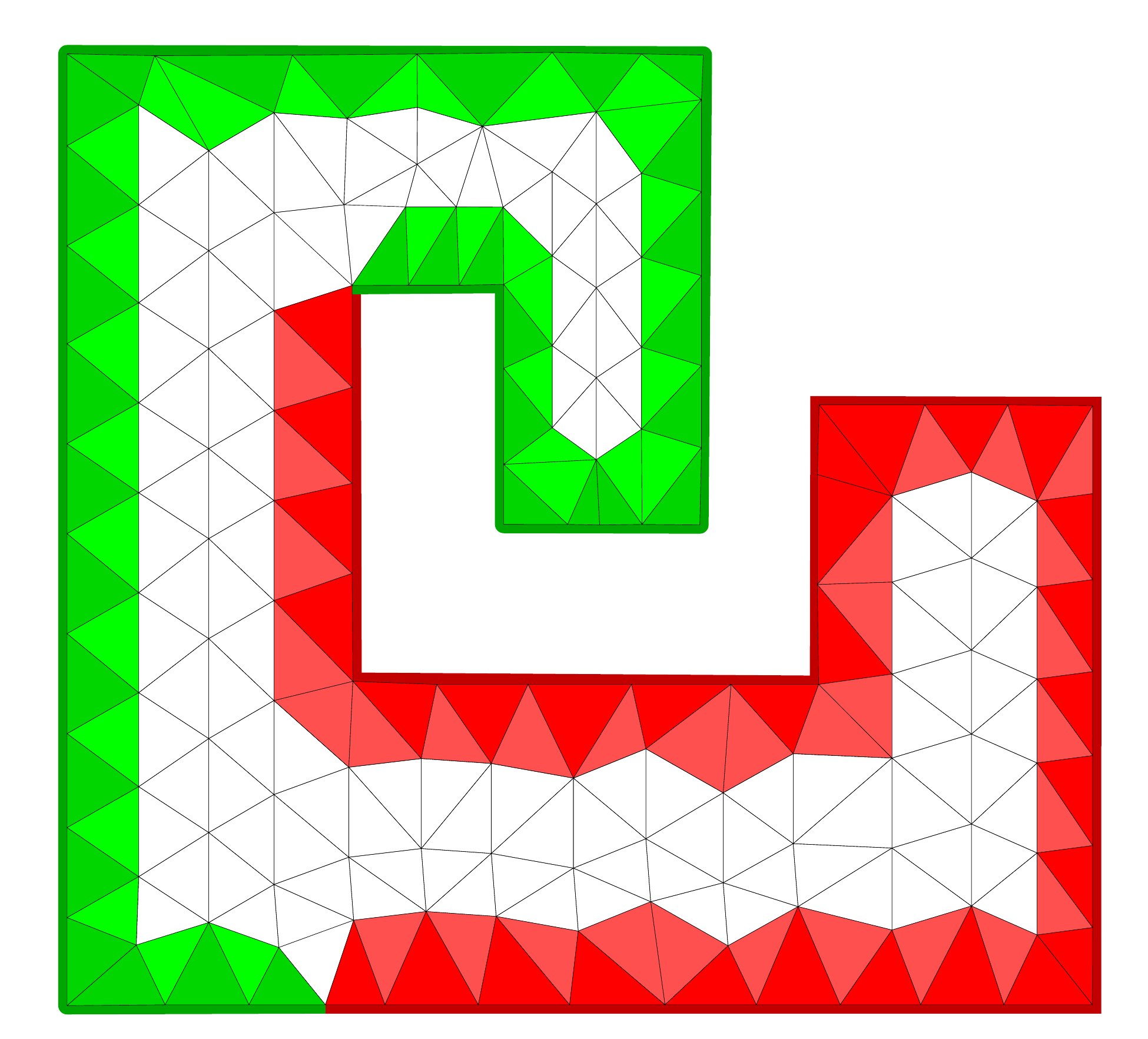}
  \end{center}
\end{wrapfigure}Given the 3D object $\mathcal{O}$ discretized using a tetrahedral mesh $\mathcal{M}$, whose surface triangles $s \in \mathcal{S}$ and their containing tetrahedra $t_s \in \mathcal{T}_s$ correspond to one of $n$ possible attributes $A_i$, we partition it by assigning each mesh tetrahedron to a unique part $\Obj_i \in \mathcal{O} $. We encode conformity by requiring each part $\Obj_i$ to be associated with a given attribute $A_i$ and to contain all surface tetrahedra $t_s$ associated with this attribute. Since we require each part to be manifold,  we constrain all tetrahedra incident on surface edges shared by pairs of triangles associated with the same attribute $A_i$ to be contained in $\Obj_i$, and similarly require any tetrahedra incident on surface vertices surrounded by triangles associated with the same attribute $A_i$ to be contained in $\Obj_i$ (see inset). We ensure the existence of a mesh that allows such labeling during our initial meshing stage (Section~\ref{sec:init_mesh}). 

We express our preference for interface smoothness and part size balance via the following  {\em partition quality} measure:
\begin{equation}
E_P =  \sum_{f \in I} \frac{A(f)}{A'} +  \omega \sum_i   \sum_{t \in \Obj_i} \frac{V(t)}{V'} d(t, \mathcal{S}_i) , 
\label{eq:energy}
\end{equation} 
Here $f \in I$ are mesh faces located on the interfaces between different parts, 
$A(f)$ is the area of face $f$, $A'$ the average mesh face area, $V(t)$ is the volume of the tetrahedron $t$, $V'$ the average volume of the mesh tetrahedra, 
$\mathcal{S}_i$ is the set of outer surface faces associated with the attribute $A_i$, and $d(t, \mathcal{S}_i) = \min_{s \in \mathcal{S}_i} |t - s|$ is the minimum distance from the centroid of $t$ to $\mathcal{S}_i$.  
The first term promotes partitions with more compact and thus smoother interfaces. 
The second term balances part sizes by penalizing partitions that associate tetrahedra far away from each given surface region with the attribute value of that region. We empirically define $\omega = 3 (\bar{A}/\bar{V})^{2/3}$ where $\bar{A}$ is the model's surface area and $\bar{V}$ is volume. This scale factor seeks to balance the terms in a resolution independent manner.

%

\begin{figure}
\includegraphics[width=\linewidth]{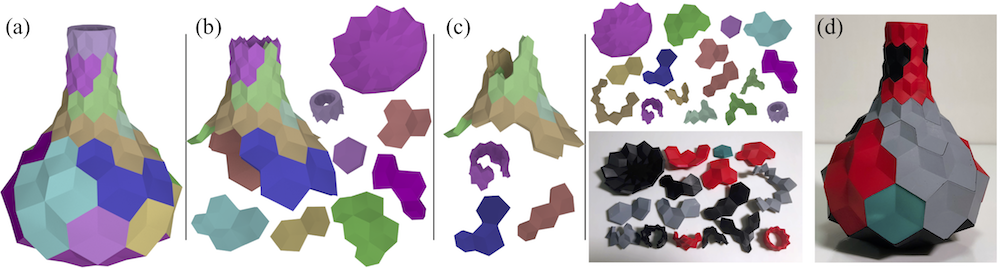}
\caption{\old{Model update}\revised{Multi-pass model disassembly process} (left to right): initial model; model after first disassembly pass; second pass; final parts and fabricated output.}
\label{fig:update}
\end{figure}

\begin{figure}
   \centering
   \includegraphics[width=\linewidth]{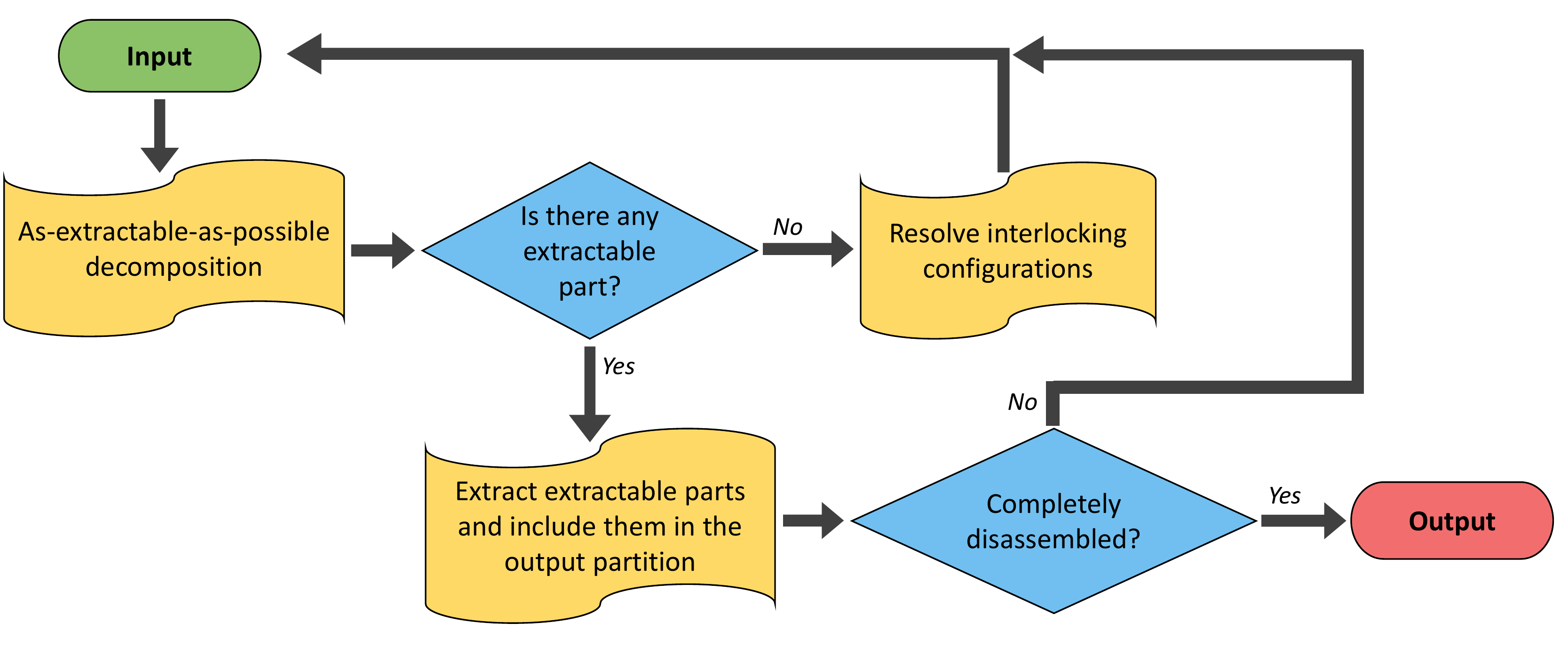}
\caption{Multi-pass disassembly algorithm.} 
\label{fig:flow_chart}
\end{figure}

\subsection{Algorithm Overview}
\label{sec:algo}

The input to our disassembly process is a closed manifold surface mesh $\mathcal{S}$ segmented into $n$ connected regions $\mathcal{S}_i$, each associated with an attribute $A_i$. Its output is a valid partition designed to satisfy our desired characteristics.  
While in general we aspire to segment the input model into parts in one go, many inputs can only be assembled or disassembled gradually, a subset of parts at a time (Figure~\ref{fig:update}). We address such models by embedding our core partition algorithm within a multi-pass disassembly process, designed to compute partitions gradually.

\paragraph{Multi-Pass Disassembly (Figure~\ref{fig:flow_chart}).} 
Each iteration of our disassembly process computes a partition of the current model that maximizes the number of simultaneously extractable parts as described below (Section~\ref{sec:aaap}). It then checks which of the resulting parts can be extracted, removes them from further consideration, and includes them in the output partition. If the model is only partially disassembled at the end of this iteration (Figure~\ref{fig:update}b) it treats the remaining solid as a new input model with a surface segmentation defined as described in Section~\ref{sec:input_update} and conducts the next disassembly iteration  on this input (Figure~\ref{fig:update}c). If at the end of an iteration none of the parts were removed, it employs one of the 
interlocking resolution strategies detailed in Section~\ref{sec:res_interlock}.  The process terminates when the model is fully disassembled. 

\subsubsection{As-Assemblable-As-Possible Partitioning}
\label{sec:aaap}

This main step of our method seeks to partition the given input into surface conforming parts, maximizing the number of simultaneously assemblable ones.
\revised{Optimizing for multiple pieces at once allows us to create various assembly alternatives, making the process easier than single sequence (i.e. puzzle-like) decompositions.
Moreover, as observed in~\cite{MLSSP18}, optimizing for a specific assembly sequence increases the risk to create too thin or too fragile pieces, because at each step the space of extractable configurations becomes smaller}.
Computing the desired partition using our underlying discretization requires solving for three sets of unknowns: the extraction directions of each part; the discrete mesh partitioning, or assignment of mesh tetrahedra to their corresponding parts;  and the geometry, or vertex positions, of the interfaces, or shared boundaries, between these parts.  
This problem contains a mixture of discrete and continuous variables, which are practically impossible to optimize for in tandem.
Surface2Volume efficiently achieves a desirable partition by uncoupling these variables and solving for one set of unknowns at a time.
 
It first identifies {\em feasible}, or \old{surface}\revised{region} extractable parts, by analyzing the input surface regions and computes initial per-part directions for these feasible parts (Section~\ref{sec:dir_init}). The computed extraction directions satisfy part \old{surface}\revised{region} extractability and maximize the likelihood of the resulting volumetric parts being extractable.  
It then computes a discrete mesh partition that seeks to maximize part \emph{interface} extractability with respect to these directions (Section \ref{sec:labeling}).
This discrete partitioning step defines the connectivity of the part interfaces and their approximate locations.  
Finally the method optimizes part extractability by modifying the geometry of the interface surfaces between them (Sect. \ref{sec:interface_smoothing}). After the discrete partitioning terminates, the topology of the interface surfaces is fully determined, allowing for interface optimization to be formulated entirely in terms of interface vertex locations. This step no longer requires the underlying tet mesh, as the parts are fully defined by the outer surface and the interfaces.   
\subsubsection{Resolving Interlocking Configurations.}
\label{sec:res_interlock}

Our partitioning step may converge to partitions in which none of the parts is assemblable, blocking further processing. 
Such interlocking will occur when the input surface segmentation does not allow for an assemblable partition in general; this is for instance the case in Figure~\ref{fig:dragon}. In this example none of the input surface regions are \old{surface}\revised{region} extractable, thus volumetric partitioning with the current region configuration is not even attempted.  When this scenario is encountered our method proceeds to segment one of the surface regions into \old{surface}\revised{region} extractable sub-regions (Section~\ref{sec:split}), associates them with distinct new attribute values, and repeats the partition algorithm on this input. 

The failure to compute any assemblable parts may also be due to our default approach, which seeks to maximize the number of parts that can be assembled in no particular order by
enforcing assemblability constraints on all feasible parts at once. While this approach works well for many inputs, it can fail on inputs which only support partitioning with highly restrictive assembly order (Figure \ref{fig:more_extractable}). Thus when as-assembable-as-possible partition fails on inputs with \old{surface}\revised{region} extractable regions, our method explores the sequential assembly option, computing one assemblable part at a time (Section~\ref{sec:input_update}).

\begin{figure*}[h]
	\includegraphics[width=\textwidth]{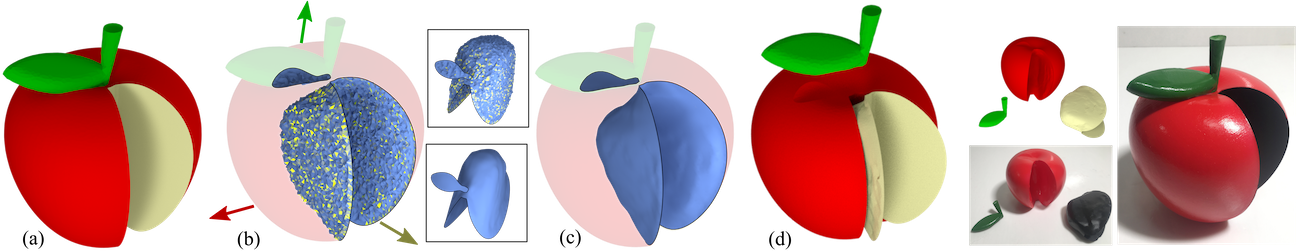}
	\caption{As-assemblable-as-possible partition: (a) input object; (b) initial directions and mesh partition interfaces (alternative view in top inset); (c) partition with optimized interfaces (alternative view in bottom inset); (d) final parts and printed object. \revised{Where mesh partition interfaces are shown, blue represents triangles that are extractable, and yellow represents triangles that are not extractable.}}
	\label{fig:aeap}
\end{figure*}

\section{As-Assemblable-As-Possible Partitioning}
\label{sec:aeap_decomp}

\subsection{Direction Initialization.} 
\label{sec:dir_init}

We compute per-region extraction directions using brute-force \old{surface} \revised{region} extractability assessment over a discrete set of possible directions. 
When no direction exists that makes the surface region $\mathcal{S}_i$ \old{surface} \revised{region} extractable, we classify the part $\Obj_i$ as infeasible. We use $I'$ to denote the set of feasible parts. 

We assess \old{surface} \revised{region} extractability with respect to a discrete dense set of directions. We use a unit-radius, uniformly triangulated sphere $\Sigma$ to represent a sampling of all the possible extraction directions, i.e., each vertex $d$ in $\Sigma$ defines a possible direction (in our implementation, $\Sigma$ has $4096$ vertices.) 
We augment this set of directions, and the corresponding sphere mesh, by including frequently seen normal directions on the input model (a normal direction is deemed to be frequently seen if it is shared by at least 1\% of the input surface triangles, measured by area) and the major axis directions.

A surface region $\mathcal{S}_i$ is \old{surface} \revised{region} extractable along a direction $d$ if all its triangles are extractable along this direction. We compute the extractability $x_{k,d}$ of a triangle $t_k \in \mathcal{S}_i$ with respect to a direction $d$  by shooting rays from the triangle's vertices along $d$ and checking whether they intersect a region $\mathcal{S}_l$, with $l \neq i$. If an intersection occurs, the triangle is obstructed and we set $x_{k,d}$ to \textit{false}, otherwise $x_{k,d}$ is \textit{true}. 
 \revised{Since we want per-region directions that facilitate volumetric partitioning, rather than only requiring each region to be 
 extractable as a zero-thickness surface, we want an epsilon-thick shell formed by offsetting the interior vertices of each region inward along the normal to be extractable along the selected directions.}
\old{In addition to requiring the vertices of each region to be extractable along each possible extraction direction, we require an epsilon-wide shell formed by offsetting the interior vertices of each region inward along the normal to be extractable.} Specifically, we offset vertices in the interior of each region by 10\% of the average edge length along the inward pointing normal, and reapply the extractability test to these offsetted vertices. We identify a region's direction as feasible if all vertices pass this test.



\setlength{\intextsep}{0pt} 
\setlength{\columnsep}{4pt} 
\begin{wrapfigure}{r}{0.25\linewidth}
  \begin{center}
    \includegraphics[width=\linewidth]{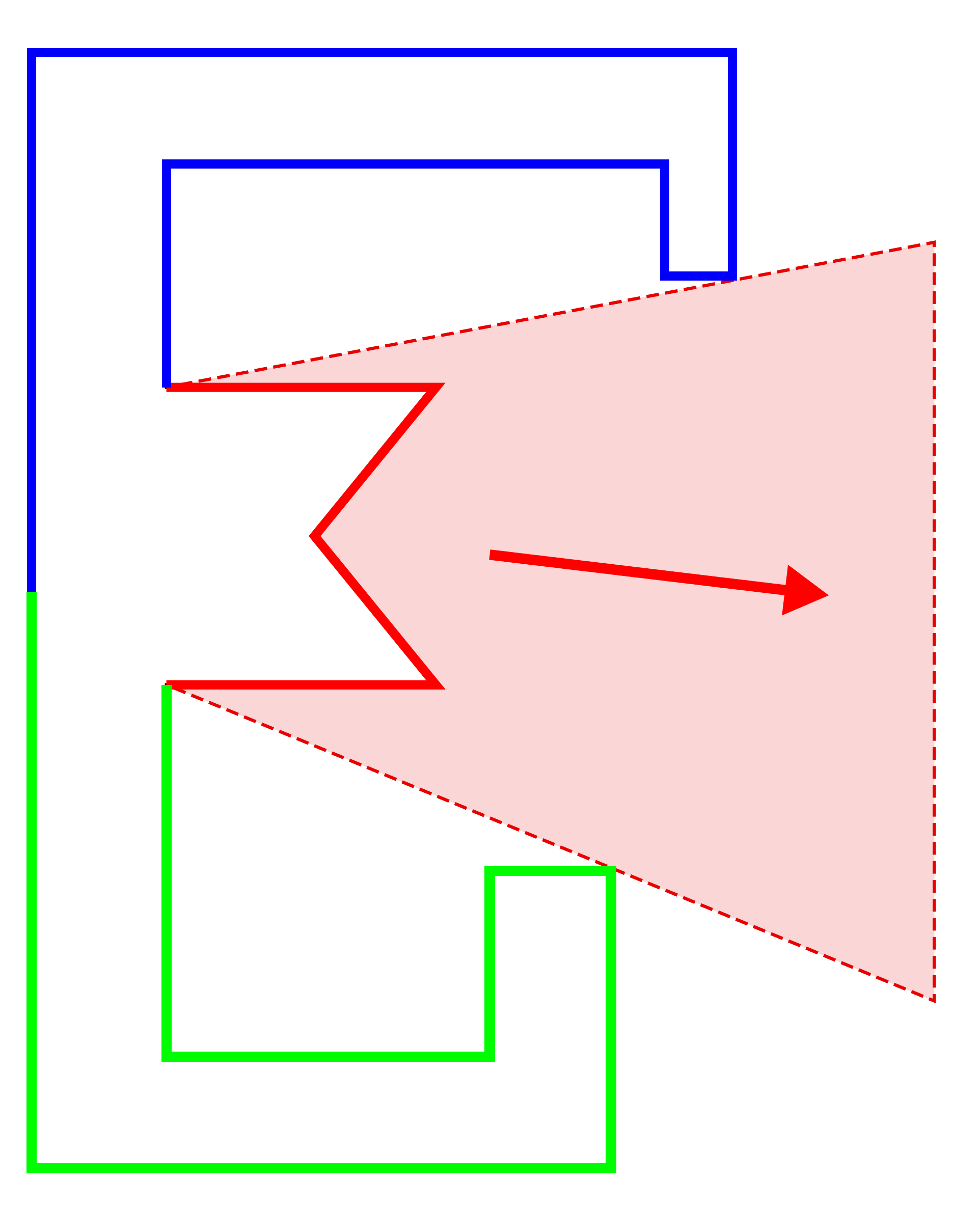}
  \end{center}
\end{wrapfigure}

Given multiple feasible extraction directions per region, we prioritize those most robust to fabrication inaccuracies and numerical errors. 
Specifically, we prioritize directions which are maximally away from the closest infeasible direction (see inset). Intuitively these directions are the least sensitive to numerical errors in the intersection test above, and are also more tolerant of subsequent fabrication errors. 
We select such most robust directions using the sphere mesh $\Sigma$.
We compute the dual mesh $\Sigma'$ of $\Sigma$, also embedded onto a unit sphere: facets in $\Sigma'$ that correspond to extractable directions in $\Sigma$ form a (possibly disconnected) region $R$. 
We select the direction furthest from the region boundaries. 

\subsection{Discrete Partitioning.}
\label{sec:labeling}
\label{sec:init_mesh}

\paragraph{Meshing and Part Initialization.}

We compute the initial uniformly sized tet mesh $\mathcal{M}$ using Delaunay tetrahedralization of $\mathcal{S}$  with volume-constrained mesh refinement \cite{tetgen}.  
When this step splits surface facets, we update the facet attributes to maintain the location of the input region boundaries.
We seek to partition the tetrahedra of this mesh into manifold parts conforming to the surface regions. To this end we require tetrahedra incident on edges or vertices in the interior of each region to be associated with this region; thus if a tetrahedron has faces, edges or vertices inside different regions, we split it separating the conflicting lower-dimension simplices. We then initialize the volumetric parts by assigning all tetrahedra incident on faces, interior edges, and interior vertices of each region to that region's part. 

\paragraph{Mesh Partition.}
We aim to partition the input mesh $\mathcal{M}$ into parts $\Obj_i$ such that the feasible parts are maximally interface extractable with respect to their respective directions $d_i$. 
Since discrete partitioning alone is unlikely to produce fully extractable parts, we reformulate interface extractability as a soft rather than hard constraint and introduce an extractability cost function that is minimized by discrete partitions which are likely to lead to extractable parts after the interface geometry optimization step (Section~\ref{sec:interface_smoothing}).


\paragraph{Extractability Cost.}
We define the extractability cost as follows. Recall that a part $\Obj_i$  is interface extractable with respect to an extraction direction $d_i$ if the outward normals along its interfaces point in the opposite direction to $d_i$. We recast this constraint as a soft cost function by integrating the amount of violation along the interface triangles $f$ that bound feasible mesh parts:
\begin{equation}
E_{E'} = \sum_{i \in I' } \sum_{f \in F_i} \frac{A(f)}{A'} \max(0,  n^i_f \cdot d_i)
\label{eq:extraction}
\end{equation}
where $F_i$ are the interface faces of part $i$  and $n^i_f$ is the outward pointing normal of $f$ with respect to part $i$. Note that this sum counts faces on the interface between feasible parts twice, once for each part. When this cost is zero all feasible parts are extractable. 
\setlength{\intextsep}{3pt} 
\setlength{\columnsep}{4pt} 
\begin{wrapfigure}{r}{0.25\linewidth}
  \begin{center}
    \includegraphics[width=\linewidth]{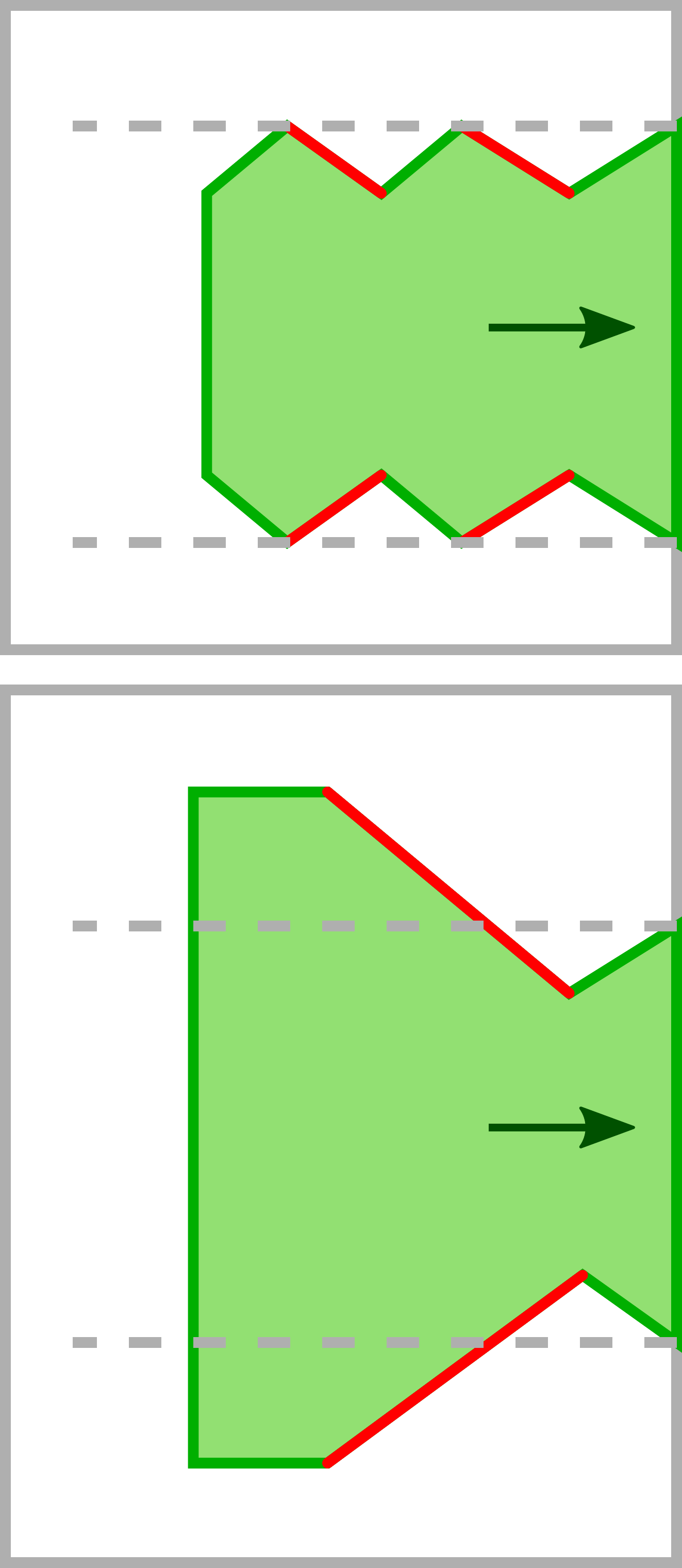}
  \end{center}
   \label{fig:why_channel}
\end{wrapfigure}
However this cost function alone is not sufficiently discriminative in distinguishing
between partitions where the violation is due to local variation in triangle normals (inset, top), and those whose interfaces are consistently misoriented (inset, bottom). 
While in the former case a local perturbation of interface vertex positions is sufficient to obtain an extractable partition, in the latter scenario making the part extractable would require significant change in vertex positions.  
We distinguish between these scenarios by leveraging an additional property of extractable parts. We note that if a part $\Obj_i$ is extractable, then it is entirely contained in the volume formed by sweeping its corresponding surface region $\mathcal{S}_i$ along the inverse of the projection direction $d_i$ (delineated by dashed lines in the insets). 
We refer to this requirement as the {\em in-channel} condition. 
This condition is satisfied in the top example, and is significantly violated in the bottom one, though the value of $E_{E'}$ is exactly the same in both cases.
We use the in-channel property to differentiate between partitions that are ``almost'' extractable and those likely to require more major changes. 
Our combined cost function $E_E$ penalizes both inextractable interface faces and faces which are outside the part channel:
\begin{equation}
E_E = \sum_{i \in I } \sum_{f \in F_i} \frac{A(f)}{A'} [\max(0,  n^i_f \cdot d_i) + IC(f,i)].
\label{eq:extraction1}
\end{equation}
Here $IC(f,i)=0$ if the face satisfies the in-channel condition with respect to part $i$, and is a constant $k_{IC}>0$ if it is does not ($k_{IC}=5$ in our implementation). To compute $IC(f,i)$ we shoot a ray from the centroid of $f$ along the direction $d_i$. If the ray intersects a surface triangle outside the region $\mathcal{S}_i$ then the triangle is marked as outside the channel (we use the centroids rather than corners to allow some leeway in the channel assessment).   
%

\paragraph{Partitioning.}
We cast our partition goal as optimizing a combined cost function which accounts for both extractability and overall partition quality:

\begin{eqnarray}
\Phi(\mathcal{M},D)  =  E_E + w_p E_P \nonumber \\
=   \sum_{i \in I' } \sum_{f \in F_i} \frac{A(f)}{A'} [\max(0,  n^i_f \cdot d_i) +  IC(f,i)]  \nonumber \\
+  w_p \Bigl(  \sum_{f \in I}\frac{A(f)}{A'} + \omega \sum_i  \sum_{t \in \Obj_i} \frac{V(t)}{V'} d(t, \mathcal{S}_i) \Bigr)
\label{eq:combined}
\end{eqnarray}

We set $w_p=0.1$, prioritizing extractability over part quality.
Computing a mesh partition that minimizes this function amounts to solving a classical multi-cut problem on the dual graph of the mesh $\mathcal{M}$, whose nodes represent tetrahedra and  whose arcs correspond to shared faces between these tetrahedra. An arc is cut if its two end nodes correspond to differently-labeled tetrahedra in $\mathcal{M}$. Finding an \emph{optimal} partition is equivalent to finding a cut that minimizes a sum of {\em unary} costs (i.e. the cost of labeling a node/tetrahedron with one of the attributes) plus a sum of {\em binary} costs (i.e. the cost of labeling an arc/adjacent tetrahedra with two of the attributes) \cite{graphcut3}.
Specifically, the function $\Phi(\mathcal{M},D)$ dictates the following {\em binary} cost of assigning labels $l_i, l_j$ to the pair of tetrahedra $t_i,t_j$ that share an internal facet $f$:

\begin{equation}
\Phi_{\textsf{Cut}}(f) = \left\lbrace
\begin{array}{cl}
0    & \text{~if~} l_i = l_j\\
\frac{A(f)}{A'} [ w_p  +  \Phi_{\textsf{Cut}}^i(f) +  \Phi_{\textsf{Cut}}^j(f)] & \text{otherwise}
\end{array}
\right.
\label{eq:phi}
\end{equation}
where
\begin{equation}
\Phi_{\textsf{Cut}}^i(f) = \left\lbrace
\begin{array}{cl}
\max(0,  n^i_f \cdot d_i) +  IC(f,i)  & \text{if~} i \in I'  \\
0 & \text{otherwise}
\end{array}
\right.
\label{eq:phi_p}
\end{equation}

The {\em unary} cost is dictated purely by our partition quality metric and is defined as  
\begin{equation}
\Phi_{\textsf{Comp}}(t,i) = w_p \omega \frac{V(t)}{V'} d(t, \mathcal{S}_p),
\label{eq:unary}
\end{equation}
for each tetrahedron $t$ and label $i$. To compute a labeling that optimizes the combined cost function, we use the \emph{gco-v3.0} multi-label optimization code \cite{graphcut1, graphcut3}: since this problem is NP-complete, the \emph{gco-v3.0} code exploits advanced heuristics to find an approximate solution in a reasonable time.

\paragraph{Enforcing Manifoldness.} 
Our subsequent interface optimization step expects the interfaces between any two adjacent parts to be two-manifolds with boundary. While
most of the time the result of the discrete partitioning as performed so far satisfies this assumption, it may occasionally contain singularities.
We eliminate such singularities by processing each interface between two of the parts independently, turning it into a manifold  if it is not one yet by duplicating vertices and edges as necessary~\cite{gueziec2001}.

\begin{figure}
 \includegraphics[width=\linewidth]{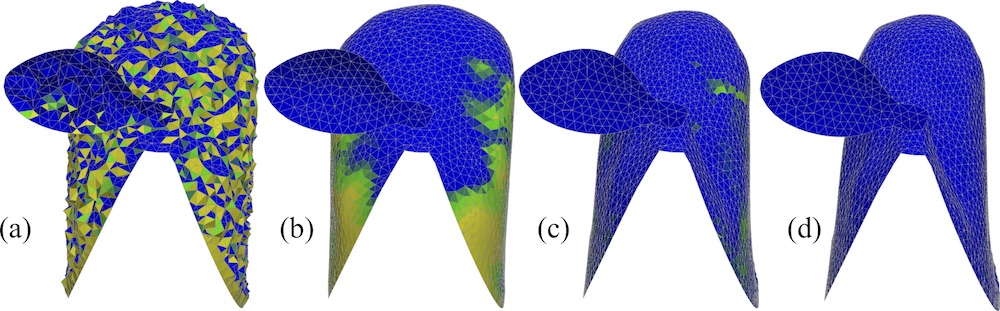}
\caption{Interface optimization for the apple in Figure ~\ref{fig:aeap} (left to right): interfaces produced by the discrete partition stage, interfaces after initial smoothing, after the first optimization iteration,  final extractable interfaces. \revised{Green areas indicate higher extractability error; blue areas are extractable.}}
\label{fig:interface_optimize}
\end{figure}

\subsection{Interface Optimization}
\label{sec:interface_smoothing}
Having found a discrete mesh partition, we proceed to optimize the geometry of the interfaces between parts. The core goal of the optimization is to strictly enforce interface extractability for all feasible parts. Its secondary goal is to smooth these interfaces to produce easier-to-manufacture parts while retaining the current parts sizes. Since the current interface locations are expected to satisfy the relative size criterion, we account for size implicitly by seeking to keep the interface vertices close to their current locations.  We formulate these goals as:
\begin{align*}
\min_{v' \in I'} E_o=  \:\: \alpha \!\! \sum_{v'_a \in I'} (v'_a - \tilde{v}_a)^2 + 
(1-\alpha) \sum_{v'_a \in I'} (v'_a -  \frac{1}{|N(a)|} \!\!\!\!\! \sum_{v'_b \in N(a)} \!\!\!  v'_b)^2\\
s.t.\\
n^i_f \cdot d_i \leq 0,  \qquad \forall i \in I', \forall f \in F_i,\\
v'_a = \tilde{v}_a \qquad \forall a \in B
\end{align*}
where $\tilde{v}_a$ are the vertex positions along the interfaces produced by our discrete partition step, $v'_a$ are the unknown interface vertex positions we seek for, $N(a)$ are the vertices immediately adjacent to vertex $a$, $n_f^i$ are the outward pointing interface facet normals defined with respect to part $i$, and $B$ is the set of vertices on the boundaries between the interfaces and the outer surface of the input model. The first component promotes fidelity with respect to the discrete partition solution. The second promotes smoother part interfaces with well-shaped triangles. The parameter $\alpha$ balances the two components and is set to $0.85$ in all our experiments. 
The inequality constraints ensure that each interface facet is extractable with respect to its associated extraction direction $d_i$, and the equalities ensure that the input surface geometry is preserved. 

While the objective function we wish to optimize is quadratic, our inequality constraints, when expressed as a function of the unknown vertex positions, are non-linear and hard to enforce using off-the-shelf approaches. 
We efficiently compute a desired solution by employing a dedicated solver that leverages the specific characteristics of our problem. We first drop the inequality constraints and compute vertex locations $v'$ that minimize our energy function $E_o$, subject only to the equality constraints (Figure~\ref{fig:interface_optimize}c); we then focus on satisfying the inequality constraints while minimally deviating from this initial less constrained solution. 

This first relaxed minimization step requires solving a simple quadratic optimization problem with linear constraints. We obtain the desired minimizer by solving the corresponding linear system using Cholesky decomposition. In theory this step could increase the number of violated per-triangle constraints. However our initial interface meshes are often very jaggy with multiple triangles significantly violating the extractability constraints. In practice, after the first minimization step which smoothes the interface surfaces, both the number of violations and the amount of violation decreases (see Figure~\ref{fig:interface_optimize}bc), making subsequent optimization more stable. 
Since the vertex positions $v'$ of our relaxed solution minimize the energy $E_o$ subject to the equality constraints, we now focus on locating the closest surface to this solution that satisfies all the constraints, including the inequalities. 

\begin{wrapfigure}{l}{0.25\linewidth}
  \begin{center}
    \includegraphics[width=\linewidth]{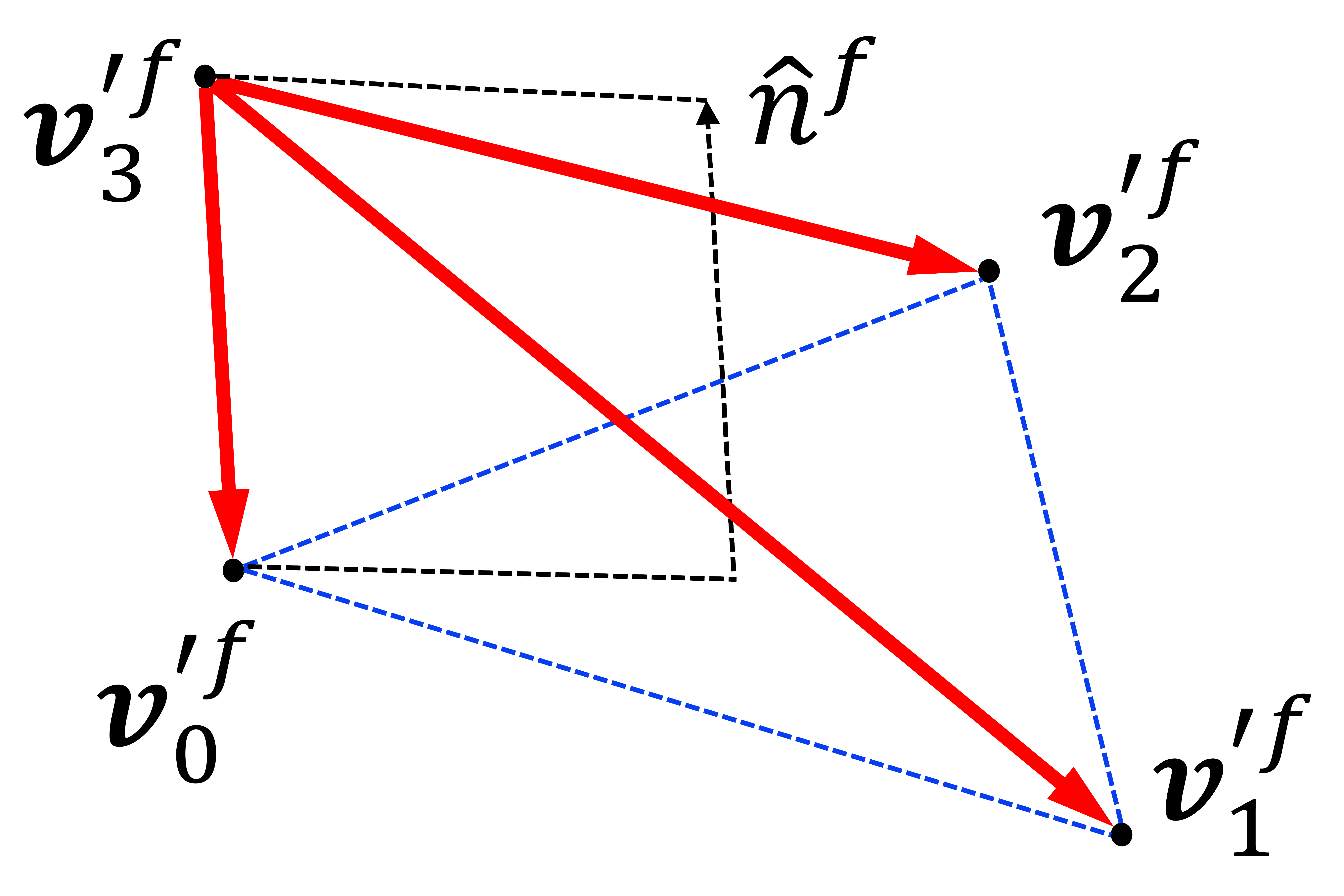}
  \end{center}
\end{wrapfigure}We formulate both the inter-surface differences and the extractability constraints in terms of per-triangle transformation gradients~\cite{sumner2004deformation}. For each interface triangle $f$ with vertices $v'^f_0,v'^f_1,v'^f_2$ we introduce a vertex $v'^f_3 = v'^f_0 + n_f$ (where $n_f$ is the normal to the triangle) and use it to define a local coordinate frame $V'_f = v'^f_0 - v'^f_3,v'^f_1 - v'^f_3, v'^f_2 - v'^f_3$ (see inset). 
We express the per-triangle coordinate frames $V_f$ of the output interface triangles in terms of the unknown output vertex locations using a similar formulation (see ~\cite{sumner2004deformation} for details). Given these two sets of coordinate frames, we express our goal of minimizing the difference between the input and output interfaces as minimizing
\[
E_c = \sum_{ i \in I',  f \in F_i} \| {V'_f}^{-1}V_f - I\|_F.
\]
where $I$ is a $3x3$ identity matrix and $\|\|_F$ is the Frobenius norm. Optimizing this energy subject to the boundary vertex placement constraints $v = \tilde{v}_a \forall a \in B$ reproduces the initial smoothed interfaces.  We incorporate the non-linear inequality constraints into this formulation using a combination of two techniques: active set optimization and linearization. 

Following a traditional active set approach ~\cite{Nocedal}, we convert the inequality constrains $n^i_f \cdot d_i \leq 0$ that encode extractability into equality constraints $n^i_f \cdot d_i = 0$ that encode normal orthogonality and selectively enforce these constraints on only a subset, or active set, of the interface triangles, those most prone to violate our inequalities. 
We initialize this active set $A$ with the triangles on the current interfaces that violate the extractability constraints. We then solve for an interface where the orthogonality constraints are enforced only for the active set:
\begin{align*}
\min_{v \in I} \sum_{ i \in I',  f \in F_i  } \| {V'_f}^{-1}V_f - I\|_F\\
s.t. \qquad n^i_f \cdot d_i = 0, \qquad \forall f \in A,  \qquad v_a= v'_a,  \qquad \forall a \in B.
\end{align*}
If any new triangles currently outside the active set violate the extractability constraints after solving the above problem, we add them to the set $A$  and repeat the process. The second technique we employ is linearization: instead of enforcing the non-linear orthogonality constraints for the triangles in the active set directly, we satisfy them gradually using an iterative local-global approach that seamlessly fits into our active-set strategy. 

\paragraph{Local Update} At each local update step we first augment the active set with any interface triangles not in the set that violate the extractability constraints with respect to one or both of their part extraction directions. Then, for each triangle in the set, we compute a minimal rotation $R^f$ such that the rotated triangle $R^f f$ satisfies our extractability constraints.  Assuming $n$ is the triangle normal and $d_1,d_2$ are the part extraction directions, we first compute a new target normal $n'$ that satisfies the orthogonality constraints by solving 
\begin{align*}
\arg \min_{n'} \: \Vert n - n' \Vert^2\\
s.t.\\
n' \cdot d_1 \leq 0\\
-n' \cdot d_2 \leq 0
\end{align*}
If the triangle is bounded by an outer surface edge, we constrain the new normal to be orthogonal to this edge. We split all triangle\revised{s} with two boundary edges to avoid over-constraining our system.   
We use the Gurobi solver~\cite{Gurobi} to compute the new normals. This step is very fast since the problems solved are very small. We do not enforce $\|n\|=1$ as this constraint makes the problem much harder. If the solver fails or returns a zero length vector, we set $n'=n$ (this may happen sporadically in earlier solver iterations). 
Given the new normal $n'$ we compute $R^f$  as the minimal rotation that aligns the current normal $n$ with $n'$.

\paragraph{Global Update}
Our global update step reconciles the local rotations computed for the active set triangles while minimally changing the interface triangle gradients overall. We formulate these requirements as follows:
\begin{align*}
\min_{v \in I} \sum_{ i \in I',  f \in F_i  } \phi_f \| {V'_f}^{-1}V_f - I\|_F +  \sum_{  f \in A } \psi_f \| {V'_f}^{-1}V_f - R^f \|_F\\
s.t. \qquad
v_i = v'_i \qquad \forall i \in B.
\end{align*}
The weights $\phi_f$  are set to $1$ for triangles outside the active set.  For triangles in the set we set $\phi_f$ and $\psi_f$ as follows to account for the degree to which we want to enforce each individual per-triangle constraint. The weights are dependent on both our confidence in the computed rotations $R^f$ and the relative amount of extractability violation of each individual triangle. Specifically, while we expect the final output normals to be roughly similar for adjacent triangles, our independently computed per-triangle rotations may produce highly divergent normals for adjacent triangles,  in \old{the} the extreme cases producing rotated adjacent triangles with opposing normals. 
To promote more consistent output normals we associate each active set triangle with a confidence weight $c^f$  computed as the average difference between the normal of the transformed triangle $R^f f$ and the normal of its transformed neighbors $R^{N_i(f)} N_i(f)$. We compute the current extractability error for each active set triangle as $e_f = \max(0,n^i_f \cdot d_i)$ and compute the intensity of its error $i_f$ as the ratio of $e_f$ and the average extractability error across all active set triangles. We consequently set $\phi_f = (1 - c_f)$ and $\psi_f = \lambda * c_f * i_f$, assigning a higher weight to the orthogonality constraints for triangles with higher confidence and higher error (we set $\lambda = 1000$). This step uses a simple quadratic minimization, enabling easy and robust computation.  

\paragraph{Reference Mesh Update.} Given a global solution, we could theoretically update the transformations applied to the active set triangles and repeatedly iterate in a manner similar to deformation frameworks such as ARAP~\cite{Sorkine}. Unlike these settings, however, our key consideration is strictly satisfying extractability, at the expense of deforming the input mesh if necessary. We therefore do not update the transformations directly and instead update the reference mesh after each iteration. We replace the reference mesh's vertex positions with those obtained from the most recent solve, setting $v'_i = v_i$ for all interface vertices. We then perform local Laplacian smoothing of vertices incident on triangles with degenerate angles or excessive Laplacian deltas. We do not move vertices if doing so increases the maximal constraint violation or would produce an intersection with the model's outer surface.


\paragraph{Termination.} The method terminates once one of the following conditions is met: all constraints are satisfied and all parts are extractable (that is, once $n \cdot d_i < \epsilon$ for all interface vertex normals with respect to all their bounded parts, we set $\epsilon=0.02$); the optimization converges (i.e. the maximal change in vertex positions drops below $10^{-5}$; or the number of iterations exceeds a fixed maximum (30 iterations in our setting). The two latter scenarios occur in our experience in the rare instances when the feasible parts cannot be made extractable along the specified directions.

\section{Model Update}
\label{sec:input_update}
\label{sec:more_extractable}

If at the end of the interface optimization step all parts are assemblable, the process terminates.
If after the optimization is complete, a subset of the parts is assemblable, these can be extracted, or removed from further processing (Figure~\ref{fig:update}b). If only one part remains, it is assemblable by default; otherwise, the union of the remaining parts defines a new model we need to partition. The outer surface of this model consists of the original surface regions of the remaining non-extractable parts and the newly exposed interfaces between these remaining parts and the extracted ones. To process this model we first must update the surface segmentation, associating the newly exposed interfaces with one of the subsequently formed parts. While the interface surfaces can theoretically be included in any of the final output parts, their current segmentation is already induced by the previously computed as-assemblable-as-possible partition, and this is suggestive of the respective surface segmentation. 
We therefore temporarily associate each exposed triangle with the label of its containing tetrahedron, and either confirm or modify this initial association using information from the adjacent surface regions. In a first pass, we iteratively grow each neighboring region $\mathcal{S}_i$ onto the exposed interface to cover all the triangles which are temporarily labeled as $\mathcal{S}_i$, whose association is confirmed. If some of the exposed triangles remain unconfirmed, we re-associate them with the closest region's attribute (using surface distance). The output of this step is a segmented surface mesh with attribute values specified for all surface triangles (Figure~\ref{fig:update}b). This approach guarantees that no new surface regions are added, whereas the initial temporary association may contain disconnected attribute \emph{islands}.
%
We rerun the Surface2Volume algorithm (Section~\ref{sec:algo}) on this new input, generating a new tet-mesh and performing the partition step from scratch (Figure~\ref{fig:update}c). The new volume meshing is necessary as we want the tet mesh to reflect the newly exposed interface boundaries. 

\begin{figure}[h]
	\includegraphics[width=\linewidth]{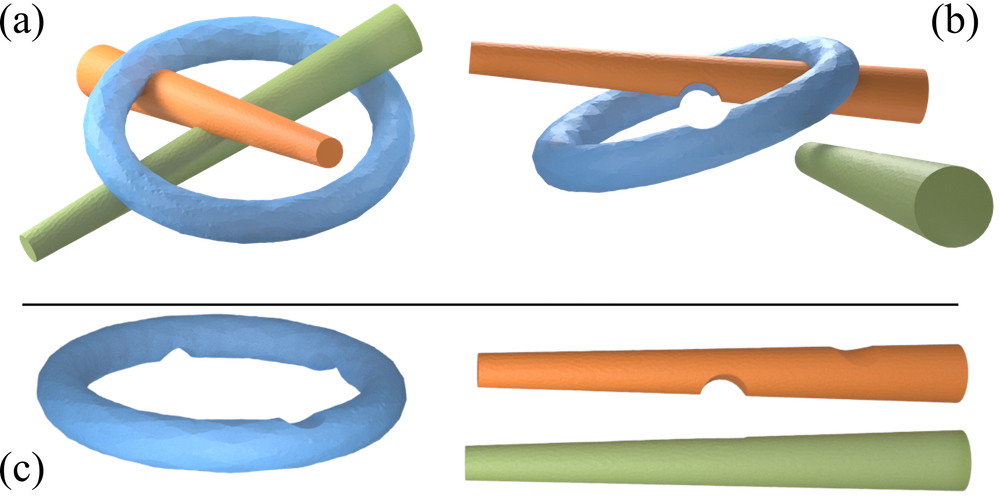}
	\caption{Sequential part extraction: (a) input with two \old{surface}\revised{region} extractable regions (orange and green) which can not produce simultaneously extractable parts; (b) first part extracted using sequential extraction (green); (c) final parts after second partition iteration.}
	\label{fig:more_extractable}
\end{figure}

\paragraph{Sequential Part Extraction}
If the method converges to a partition with no extractable parts (Figure \ref{fig:more_extractable}(a)), rather than trying to compute assemblable parts for all feasible surface regions at once, we explore the option of extracting one part at a time. 
To facilitate such sequential extraction we locate a currently non-assemblable, feasible part that is most likely to become assemblable with minimal changes. To find such a part we order all feasible parts based on the percentage of their interface area that is not extractable after optimization, and process them in increasing order of non-extractable area, seeking to locate one that can be extracted in isolation with minimal changes to its geometry. 

For each examined part we compute a locally best partition by marking this part as feasible and all others as not feasible, and re-applying the partition algorithm in Section~\ref{sec:aeap_decomp}. In this setting, the partition criteria used are dominated by the extractability constraints with respect to this part only, maximizing the likelyhood of producing an extractable part. Once we obtain a part that is extractable (Figure \ref{fig:more_extractable}(b)), we remove it  and repeat the partition process on the remaining model as described above (Figure \ref{fig:more_extractable}(c)).



%
%
%
%




\section{Segmentation Refinement}
\label{sec:splitting}
\label{sec:split}

If all surface regions are infeasible (all surface regions have no valid extraction directions) we segment the largest region, replacing it by feasible sub-regions and introducing corresponding new ``dummy'' attributes. 
We then rerun the partition on this refined input.  


Given an infeasible region, our goal is to segment it into assemblable contiguous sub-regions of roughly equal size with compact in-between boundaries.
When assessing sub-region assemblability, it is straightforward to test if the sub-region's motion is obstructed by an existing region; however, explicit assessment of whether one sub-region obstructs another sub-region before they are fully formed is not possible. Instead, while forming sub-regions we reduce the likelihood of one sub-region blocking another by growing sub-regions with maximally compact shared boundaries and far away extraction directions. 
 We start by attempting to segment the input region into just two such sub-regions, and increase the number 
of produced sub-regions if this fails. 

\subsection{Binary Segmentation}
The binary segmentation algorithm aims to split a region $\mathcal{S}$ into two new \emph{surface-extractable} contiguous sub-regions $\mathcal{S}^0, \mathcal{S}^1$.
To this end it starts by locating pairs of potential extraction directions $d_0, d_1$ and uses the pair deemed best in terms of the criteria below to compute the sub-regions. 
If the produced regions are not contiguous it proceeds to the next best pair.  

\paragraph{Extraction Directions.}
Denoting with $C_i$ the set of triangles that are extractable along direction $d_i$, a split is valid only if $C_0\cup C_1 = \mathcal{S}$ (i.e all triangles in $\mathcal{S}$ are extractable along either $d_0$ or $d_1$).
We rank the set of all valid direction pairs $\Sigma$ using the following cost function
\begin{equation}
D_{01} = \frac{1 - d_0 \cdot d_1}{2} + \alpha (1 - \frac{|\mathcal{A}_0 - \mathcal{A}_1|}{\mathcal{A}'}).
\label{eq:pair_selection}
\end{equation}
Here $\mathcal{A}_i$  is the sum of areas of the triangles in $C_i$, and $\mathcal{A}'$ is the average area difference across all direction pairs in $\Sigma$.
The first term promotes opposite extraction directions, reducing the likelihood of obstruction between the resulting sub-regions. The second term balances sub-region sizes, promoting more even output sub-region sizing.
The scalar $\alpha$ balances the two terms, and was set to $0.1$ in all our experiments. 

\paragraph{Region Growth.}
Given a pair of extraction directions $d_0, d_1$, we grow the sub-regions $\mathcal{S}^0, \mathcal{S}^1$ starting from a seed triangle and gradually adding triangles that share edges with one of the current sub-regions to this sub-region based on a priority metric. 
To determine seed triangles for the sub-regions $\mathcal{S}^i$, we project each triangle in the source region onto the ray whose origin is the region's centroid and whose direction is the extraction direction $d_i$. We then select the triangle from the source region that is furthest along the ray as the new seed for the subregion.
We use the following metric to prioritize the next triangle to add to one of the sub-regions and only add triangles to $\mathcal{S}^i$ if they are extractable along $d_i$:
\begin{equation}
\dfrac{D(t, d_j)}{D'} + \gamma \dfrac{l_{outer}(S_i^j, t)}{l_{inner}(S_i^j, t)}
\label{eq:assignment_cost}
\end{equation}
where $D(t, d_i)$ is the distance from the triangle $t$ to the ray whose origin is the centroid of $\mathcal{S}$ and whose direction is $d_i$, $D'$ is the average distance to this ray for all triangles in the region, and $l_{inner}$ and $l_{outer}$ are the overall length of $t$'s edges that are shared with the current region $S_i$ and those that are not, respectively.
The first term promotes compactness by penalizing triangles that are far away from their region's seed; the second term promotes the creation of shorter boundaries \cite{julius2005d}. We use $\gamma=0.15$.

\begin{figure}[h]
	\includegraphics[width=\linewidth]{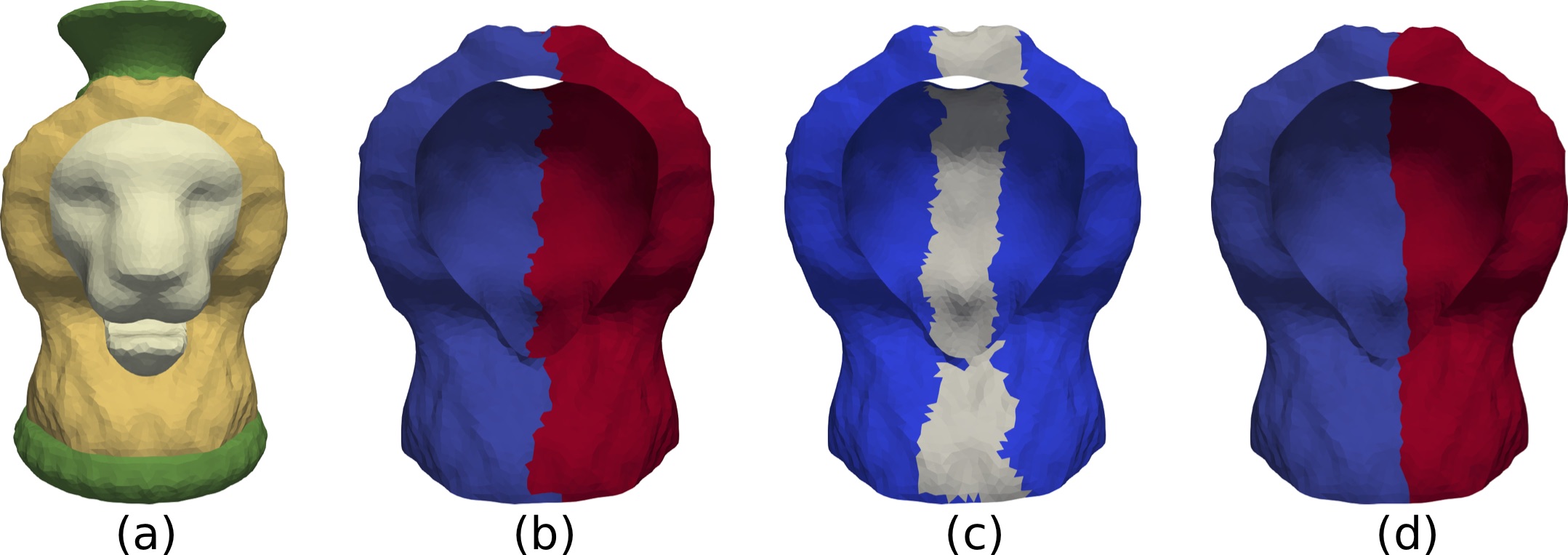}
\caption{Split boundary optimization: (a) input model; (b) region with raw cut; (c) re-assignable triangles; (d) final smooth cut.}
	\label{fig:split_opt}
\end{figure}

\paragraph{Boundary Optimization.} 
The segmentation produced by the region growing algorithm has boundaries that are linked to the tessellation, and are likely to be jagged in areas where the assignment to both regions have similar costs (Figure~\ref{fig:split_opt}~(b)). To improve the interface between adjacent pieces we optimize segmentation boundaries by isolating a strip of triangles that are close to the original one, and can be extracted along both directions (Figure~\ref{fig:split_opt}~(c)). We compute the final boundary within this strip using the level set smoothing method proposed in~\cite{Liv17_extended}, splitting edges to embed it into the connectivity of the mesh (Figure~\ref{fig:split_opt}~(d)). 
%
%

We explicitly test that neither sub-region blocks the assemblability of the other along the chosen directions. If the test fails, we proceed to regrow regions using the next best direction pair. We limit the testing to 10 pairs, and use a multi-region split if all fail.

\subsection{Segmenting into Multiple Sub-Regions. } 
If the binary split fails to produce two assemblable sub-regions, rather than splitting those recursively we repeat the same partition process but with $n$ directions and sub-regions, first using $n=3$ and increasing it until the produced sub-regions are extractable.  We select $n$ directions that minimize $\sum_{ij} D_{ij}$ (Equation \ref{eq:pair_selection}) and then proceed to seed and grow charts as before. 

\section{Multi-Region Parts}
\label{sec:unify_components}

Users often want disjoint regions to have the same attribute value - for instance, they may want to use the same color for all the legs of a caterpillar
(Figure~\ref{fig:split}).  Since they often seek to reduce part count, users may wish to keep these same-attribute regions together as outer surfaces of a common volumetric part.
Our framework supports formation of  such common parts by minimally changing the partition algorithm in the presence of such disjoint same-attribute regions.
We leave it to the user to indicate which same attribute regions should be joined into such multi-region parts.

\begin{figure*}[h]
\includegraphics[width=\textwidth]{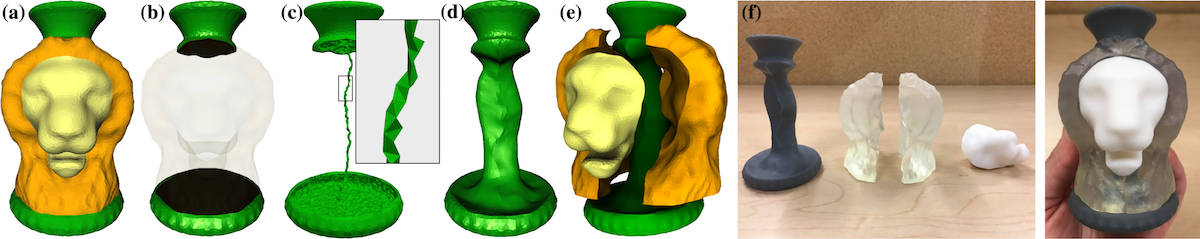}
\caption{Multi region part processing: (a) input surface; (b) two regions having the same \emph{green} attribute; (c) tetrahedra affiliated with the green attribute after MST calculation; (d) single part formed out of the two merged green regions; (e) whole object decomposition; (f) printed parts and assembled object.}
\label{fig:multi_region}
\end{figure*}

\begin{figure*}
\includegraphics[width=\linewidth]{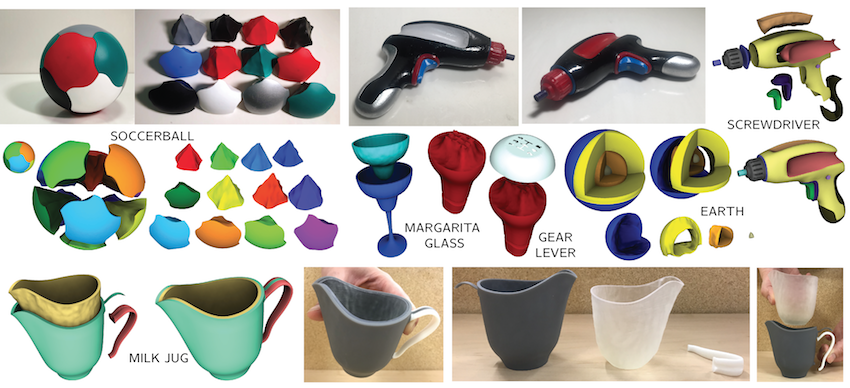}
\caption{A gallery of both digital and fabricated results obtained with our method.}
\label{fig:mosaic}
\end{figure*}

Given a set of regions users wish to keep together (Figure~\ref{fig:multi_region}(b)), we change the initial mesh element labeling (Section \ref{sec:init_mesh}) to form a single connected component for each such compound region as follows.
We first label the tetrahedra adjacent to the surface using the same process as before. We then compute the dual graph $\mathcal{D}$ of the tetrahedralization, where nodes correspond to tetrahedra (and inherit their attributes, if any) and where arcs correspond to pairs of tetrahedra sharing common facets. Each node is placed at the barycenter of the corresponding tetrahedron. Each region $\mathcal{S}_i$ corresponds to a connected subgraph in $\mathcal{D}$ where all the nodes inherit the region's attribute value. We mark nodes that do not correspond to any region as \emph{unaffiliated}. 
For each attribute $A_j$ of interest we compute a minimum spanning tree over $\mathcal{D}$ connecting all its regions along paths containing only unaffiliated nodes.
In computing the tree we seek paths which are both short and which lie away from other surface regions (so as to minimally block subsequent growth of other parts). 
 
For each region $\mathcal{S}_i$ we compute a center point $p_j$ which is geodesically farthest from the region's boundaries, then mark all tetrahedra which are at most half the distance from $p_j$ to the boundary as potential tree nodes. 
For each attribute $A_j$ we compute a minimum spanning tree over $\mathcal{D}$ connecting potential tree nodes, one for each region, along paths containing only unaffiliated nodes. We set the weight of each arc  $a$  connecting the nodes $n_1$ and $n_2$ to be its length times the inverse of its distance from the surface of $\Obj$:

\begin{equation}
w(a) = \lvert a \rvert \cdot \frac{d(n_1, n_2)}{d(n_1, \mathcal{S}) + d(n_2, \mathcal{S})}
\label{eq:mst_cost}
\end{equation} 

where $\mathcal{S}$ is the boundary of $\Obj$, and $d(.,.)$ is the Euclidean distance. This weighting moves paths inward and allows adjacent parts more freedom to grow. 
We then associate all the tetrahedra (nodes) in the computed tree with the attribute $A_j$ (Figure~\ref{fig:multi_region}(c)). 
We process attributes in descending order of the number of regions they are affiliated with; for attributes with the same region count we process the ones with smaller corresponding region area first. The latter preference decreases the likelihood of forming tiny parts. 

The rest of the partitioning algorithm is performed as described in Section~\ref{sec:algo}\old{,}\revised{ and } ~\ref{sec:aeap_decomp} with the following minor difference in \old{surface}\revised{region} extractability assessment (Section \ref{sec:dir_init}). 
For all regions we test for ray intersections not only against the other attribute regions, but also against the surface of the tetrahedra paths connecting these regions.  
For multi-region parts, we apply the test to both region and path vertices (vertices of tetrahedra traversed by the path).

\section{Results and Validation}
\label{sec:results}

\begin{figure*}
\includegraphics[width=\linewidth]{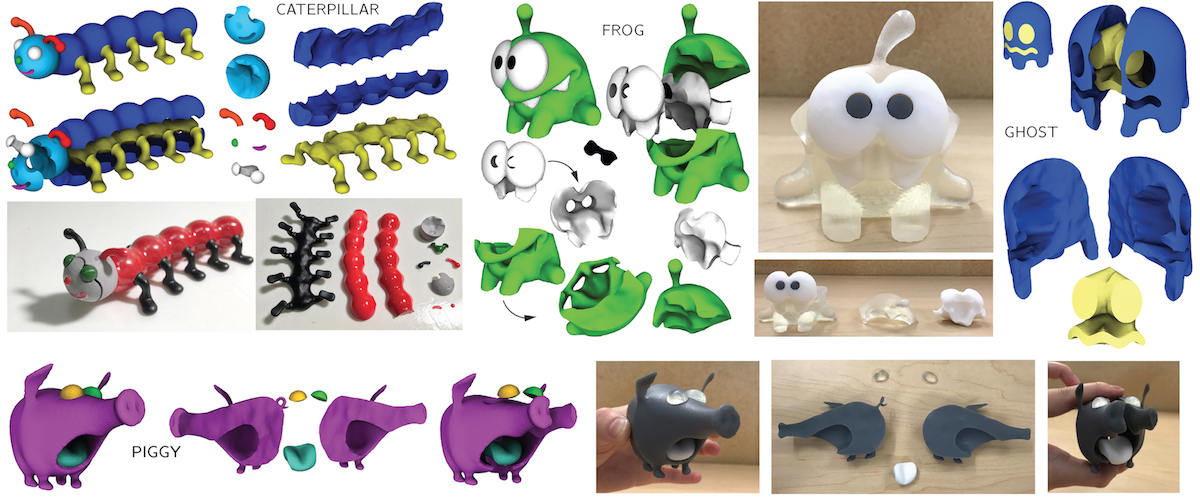}
\caption{Example inputs that require surface segmentation refinement to enable partition and our outputs. The body and the tongue of the pig model are not \old{surface}\revised{region} extractable; after splitting the body, all parts become extractable. The frog, caterpillar and ghost have small features (tooth, eyeballs and pupils for the frog, antennae and legs for caterpillar, and mouth and eyes for the ghost) that we seek to keep together when possible. These choices can only be satisfied by refining the surrounding regions.
 }
\label{split}
\label{fig:split}
\end{figure*}

\begin{figure}
\includegraphics[width=\linewidth]{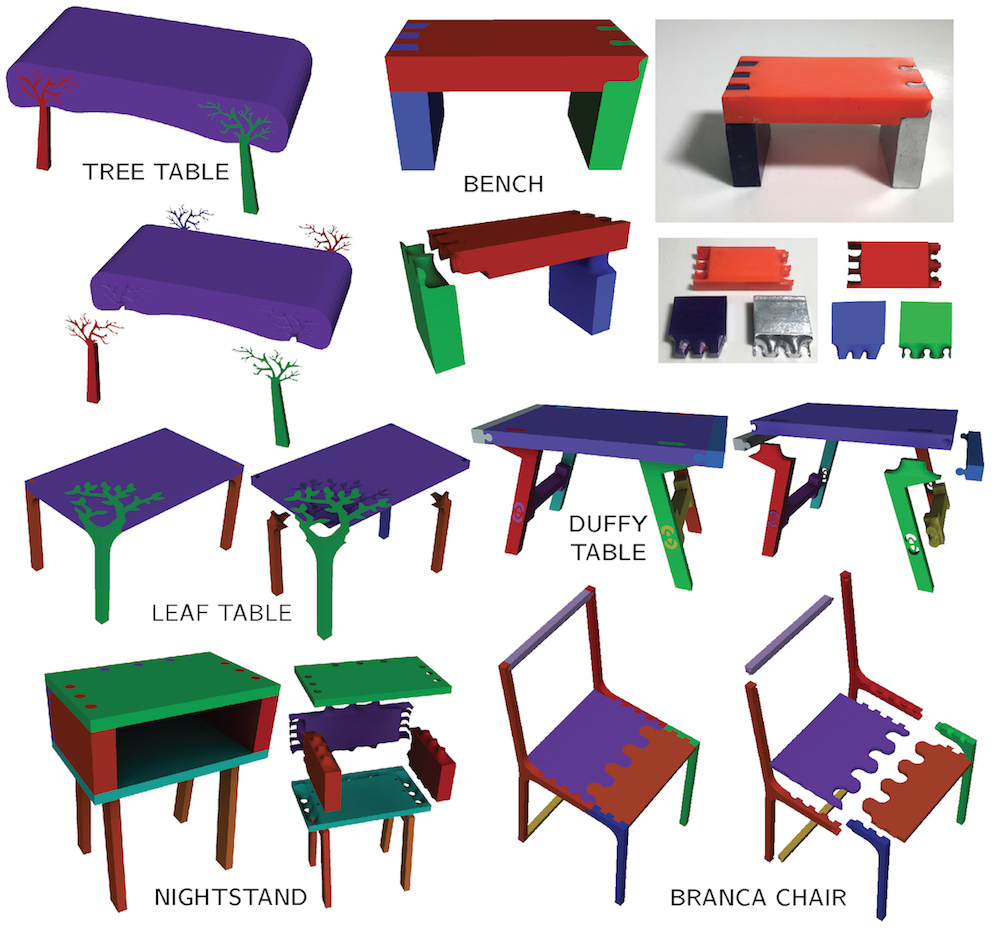}
\caption{Partitions of furniture models from Yao et al. \cite{yao2017interactive}. Our method found non-obvious directions (validated computationally and via printing) for a few parts, such as the right side of the bench and the side strips on the Duffy table.}
\label{fig:furniture}
\end{figure}

Throughout the paper we demonstrate Surface2Volume's performance on twenty-one diverse inputs. \revised{Input segmentations came from pre-existing colorized inputs, existing meshes colorized using a simple user interface, and challenging 3D models created by an artist based on images of real puzzles.}  In our tests we focused on the types of free-form objects users are likely to fabricate, including furniture (tables, chairs, bench, and the nightstand in Figure~\ref{fig:furniture}), natural shapes (caterpillar, frog, ghost, Figure~\ref{fig:split}), decorative objects (vase puzzle, apple), and engineered shapes (screwdriver, gear, Figure~\ref{fig:mosaic}). We tested both relatively smooth inputs (soccer and swirl balls) and ones with multiple fine details (lion statuette), and both genus zero and high-genus shapes (puzzle, Figure~\ref{fig:more_extractable}). Due to the effort required to print and assemble multi-part shapes, we anticipate that users would like to keep the number of different surface attributes specified to under a dozen, motivating us to focus our tests on such examples. 

Our set of inputs validates all the core elements of our method: as-assemblable-as-possible partitioning, sequential part extraction, region splitting, and multi-region attribute processing. 
For many of our inputs (for instance, the soccer ball, the swirl-ball, and the milk jug) we were able to compute parts that allow for any assembly order, and are extractable within a single disassembly pass. Others require a multi-stage assembly process, such as the Duffy table and puzzle vase (Figure~\ref{fig:update}). Our method robustly handles both scenarios, computing parts over multiple disassembly iterations for the latter scenario. It also seamlessly handles cases where segmentation conforming partitioning requires sequential part extraction, such as the puzzle in Figure~\ref{fig:more_extractable}. For the dragon and pig, the input set of regions does not allow for region-split-free partitioning.  
For others such as the lion, caterpillar, frog, and ghost, the preference for  keeping regions with similar attribute values together similarly necessitated segmentation refinement.
In all these cases our method automatically computed region segmentations and subsequent partitions that allow for subsequent part assembly. 
   

%
%

\paragraph{Output Fabrication.} 
Across the paper we exhibit a variety of 3D printed results of different complexity. In all cases we were able to assemble the fabricated parts to form the target object. To account for the inherent inaccuracy of 3D printing, we offsetted thicker parts inward along the normal direction by a small epsilon reflective of the printer tolerance. Several of these examples include the use of semi-transparent materials (wavy cylinder, milk-jug, frog) - these looks cannot be achieved by external surface painting or computational hydrographics.

\begin{figure}[h]
\includegraphics[width=.9\linewidth]{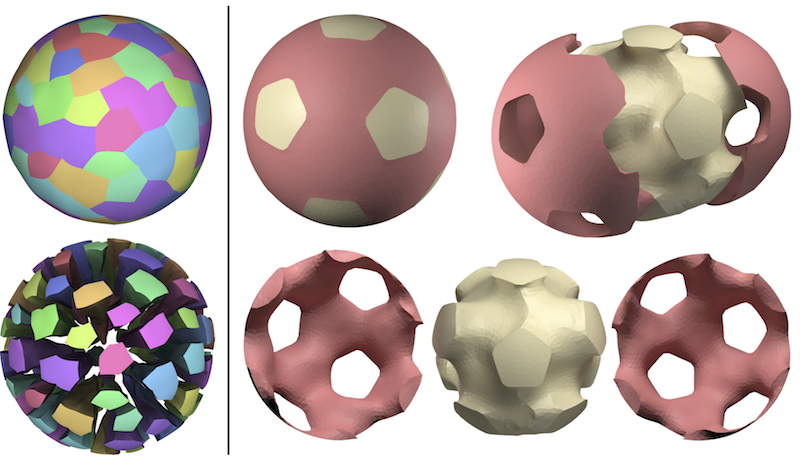}
\caption{\revised{Stress tests. Left: our method can robustly handle surfaces with approximately a hundred different materials, such as this sphere. Right: our method robustly handles surfaces with multiple disjoint regions composed of the same material, such as this soccer ball.}}
\label{soccerball}
\label{fig:soccerball}
\end{figure}


\paragraph{Comparison to Prior Art.}
As discussed in Section~\ref{sec:related}, while assemblability is a critical requirement when partitioning models for fabrication, there is little prior work on surface segmentation conforming assemblable partitioning. Basic methods that offset the segment boundaries either along a fixed direction or along the inverse of the surface normal fail on even medium complexity inputs, such as the apple or the wavy cylinder. 
At our request, Yao et al. \shortcite{yao2017interactive} attempted to partition the milk-jug, the soccer ball, the swirl-ball, and the wavy cylinder. The method failed to generate parts for the first two, generated an invalid result for the swirl-ball (Figure~\ref{fig:swirl_ball}), and was able to partition the wavy cylinder
(Figure~\ref{fig:hybrid_fabrication}) after remeshing it. Our method successfully processed these inputs algorithmically. This
\begin{wrapfigure}{l}{0.21\linewidth}
  \begin{center}
   \includegraphics[width=\linewidth]{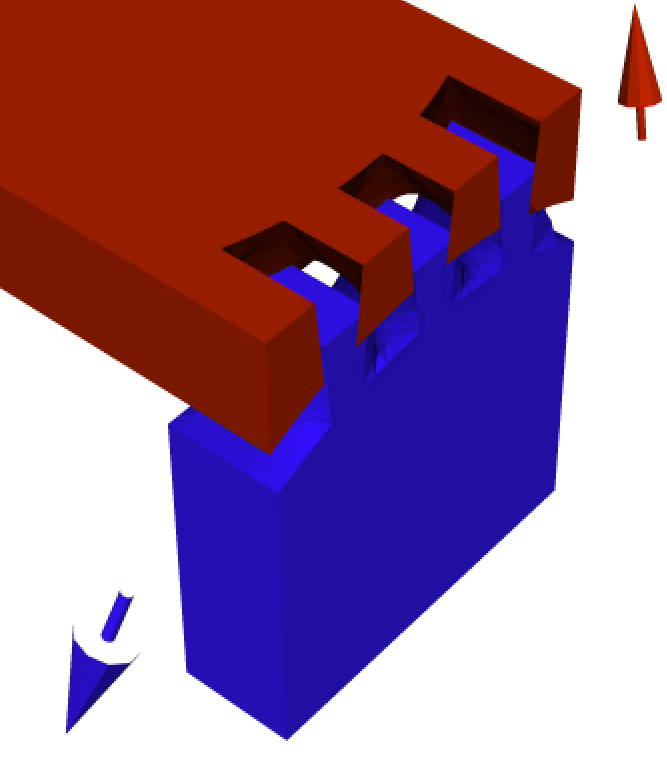}
    \includegraphics[width=\linewidth]{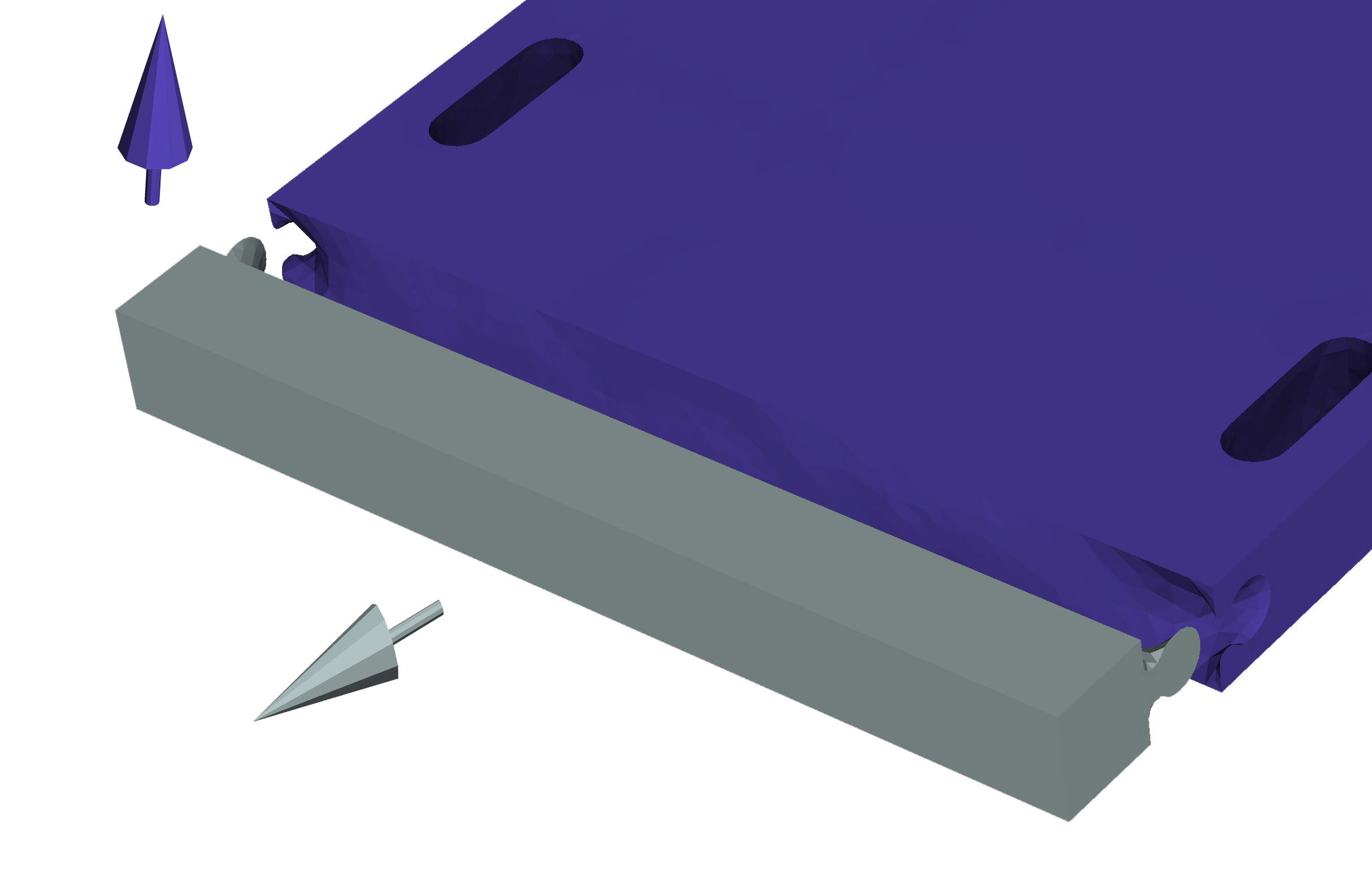}
  \end{center}
\end{wrapfigure}
comparison highlights the main advantage of our method, namely its ability to process free-form inputs that require computation of interfaces with complex, irregular interface topology and geometry. 
We tested our method on the furniture models provided by Yao et al., and our method was able to \old{process} \revised{correctly partition} most of these inputs (Figure~\ref{fig:furniture}). We discuss the failure cases in Section~\ref{sec:concl}. 
One of the most intriguing aspects of our results was that the extraction directions for a few parts found by our method were not the \revised{axis-aligned} ones that a human observer would predict (see inset). 
\revised{Specifically for the bench, Surface2Volume extracts the side part downwards and to the left, and extracts the table's side along a forward-left diagonal.} Both computation and fabrication confirmed that these directions indeed enable assemblable partitioning.

\paragraph{Parameters. \old{and User Control.}} 
Unless \revised{explicitly} stated otherwise, all of the examples shown \revised{in the paper} were generated with a single set of parameters, set as described in Sections~\ref{sec:overview} through ~\ref{sec:unify_components}. \revised{We performed a sensitivity analysis in which we ran Surface2Volume on a series of challenging models (apple, dragon, gearleaver, and the teaser model), doubling and having each of the parameters $\omega$, $w_{p}$, $\alpha$, $\varepsilon$, and $\gamma$ respectively. In all cases, Surface2Volume successfully produced an extractable partitioning of every input with the same part count as before.}

\begin{wrapfigure}{l}{0.25\linewidth}
  \begin{center}
    \includegraphics[width=\linewidth]{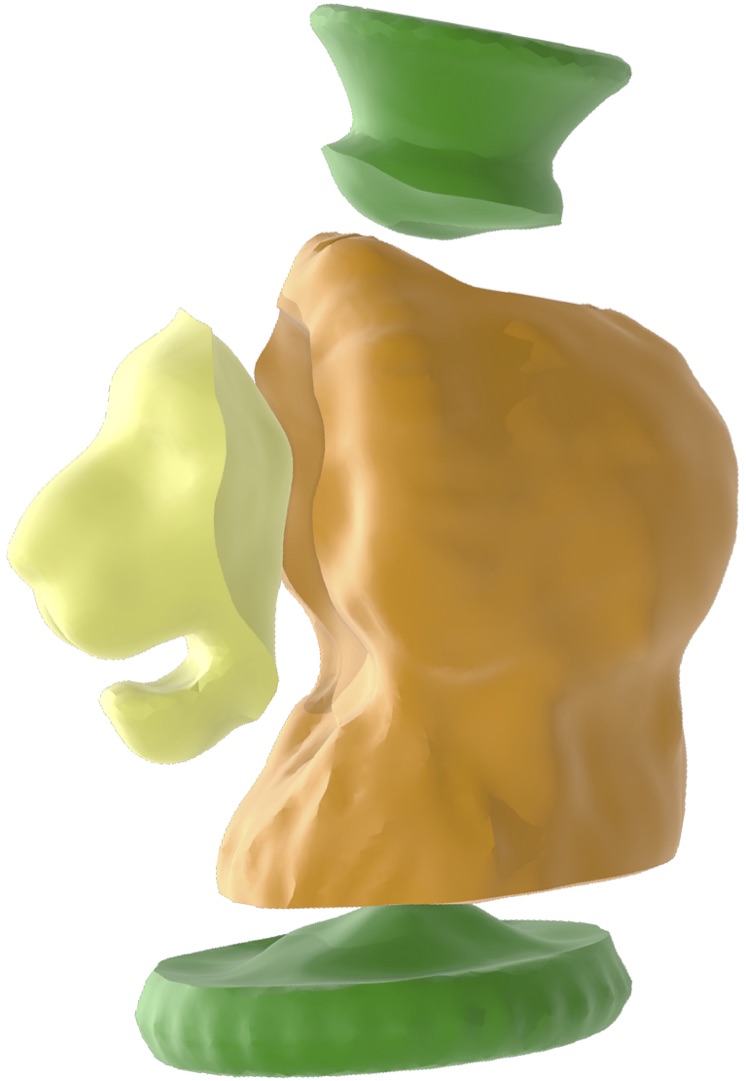}
  \end{center}
\end{wrapfigure}

\paragraph{Multi-Region Parts.}
Choosing which same-attribute regions should or should not bound common parts is inherently dependent on the user desired part-size balance.
By default we merge all regions with the same attribute \revised{value} into a single part, but leave it up to the user to state otherwise. The inset shows an alternative partitioning of the model in Figure~\ref{fig:multi_region} where this possibility was exploited. 


\paragraph{Part Extension}
\revised{The size of the parts we compute is controlled by the weight $\omega$ (Equation \ref{eq:unary}). For all inputs we can obtain extractable parts using the default setting of this weight. We enable users to control the size of individual parts by modifying this weight, enabling them to avoid 
forming parts that are too thin to be manufacturable for a given overall object size and material. This also enables accounting for material cost, by reducing the size of parts requiring expensive materials.}
\old{As noted earlier, larger parts are easier to fabricate and manipulate during assembly.
 Users can increase the size of parts if deemed too 
small by decreasing the weight $\omega$  in the unary term (Equation \ref{eq:unary}) of our graph-cut computation, for these specific parts.}
We \old{use} \revised{demonstrate} this option \old{for} \revised{on} the tree and leaf tables (Figure~\ref{fig:furniture}) decreasing \old{this value} \revised{$\omega$} by a factor of 5 for the parts associated with the narrow branch regions. \revised{Using the default values on these inputs produces valid partitions but ones with parts which are too shallow for practical purposes.}

\paragraph{Robustness.} \revised{In order to test the stability of our method on challenging inputs, we partitioned a sphere with 100 different surface colors, and a soccer ball with 12 spots where we required all spots to belong to the same part. Our method successfully produces extractable partitions for both of these inputs (Fig. ~\ref{fig:soccerball}).}

\paragraph{Runtimes and Statistics.} The input models we tested  have between \textasciitilde 20,000  and \textasciitilde 100,000  surface triangles, with the initial tetrahedralizations ranging in size from \textasciitilde 200,000 to \textasciitilde 4,000,000 elements.  In all cases our method automatically computed an assemblable partition within 10 minutes, with the exception of the models from Yao et al. \shortcite{yao2017interactive}, which took up to 2 hours due to requiring a higher resolution tetrahedral mesh to capture the fine detail. We note that Yao et al. report runtimes of 24 hours for the bench model, which we complete in under 2 minutes. \revised{Experiments were run on an AMD Threadripper 1950X CPU, with 16GB RAM, running Windows 10.}
 
\paragraph{Assembly Stability.} In our work we focus on assemblable part computation and generate parts that fit snugly together. In many cases the friction between the parts is sufficient to hold them together, while in others parts may require the use of glue to be held together. An appealing alternative, which could be investigated in the future is to add connectors to the processed parts while preserving assemblability, following on ideas presented by~\cite{AutoConnect}. \revised{Alternatively, given our parts, one can constrain part motion via a post-process that computes an extraction order and extrudes joints from each part along its extraction direction into adjacent parts that come later in this order.}

\section{Conclusions.}
\label{sec:concl}
We have presented Surface2Volume, a new method for computing assemblable segmentation-conforming partitions of 3D models. Our method outperforms prior approaches in terms of of the range of inputs it can handle, its robustness, and its efficiency. This method has immediate applications for digital fabrication, and can significantly simplify the fabrication of high quality multi-material, multi-color objects. At the core of our method is a combined discrete-continuous optimization process that uses a discrete mesh labeling approach to efficiently obtain an approximate solution for the problem at hand, and a geometric optimization process that converts this discrete solution into our desired final partition. 


\paragraph{Limitations} 

\revised{Surface2Volume operates on discrete tetrahedral inputs and is inherently limited by both the size and quality of input surface and volume mesh elements. In particular,} our method is inherently constrained by the resolution of the volumetric mesh we use 
to discretize the input model. \revised{Accordingly, Surface2Volume} can only approximate extrusion surfaces to the accuracy allowed by the mesh resolution. It successfully partitions inputs such as the puzzle in Figure ~\ref{fig:more_extractable} where the extruded parts 
are large compared to the overall model size. However, models such as the bookshelf 
\begin{wrapfigure}{l}{0.25\linewidth}
  \begin{center}
  \includegraphics[width=\linewidth]{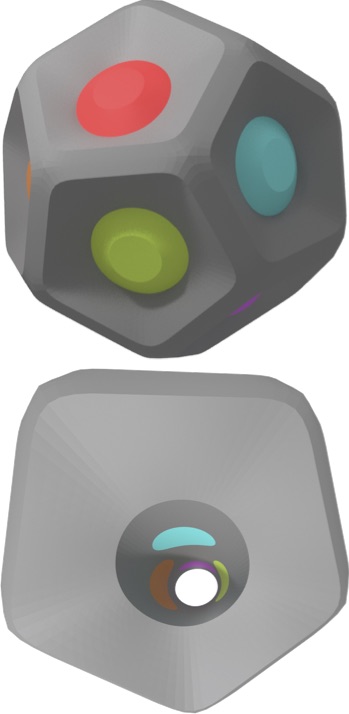}
  \end{center}
\end{wrapfigure}
and the archchair in ~\cite{yao2017interactive} require extrusion of curved segmentation boundaries with edges whose length is below 0.5\% of the model's bounding box (a tet mesh at this 
resolution would have over 100 million elements). We consequently fail to produce valid results on such inputs. \revised{Additionally, mesh quality along newly exposed interfaces (Section \ref{sec:input_update}) is not optimized after part extraction; accordingly, our method may 
become less robust on models that require multiple partition iterations. This could be solved, in practice, by remeshing the newly exposed interfaces at each update step.}
Our current implementation \revised{also} assumes that, after a subset of parts is extracted, the newly exposed interface can be associated with one of the adjacent regions (Section~\ref{sec:input_update}). This assumption simplifies our implementation, and holds for the 
vast majority of inputs we tested. It does not hold for pathological models such as the wooden puzzle (inset). As the fish-eye view in the inset (bottom) shows, after removing the puzzle's lock bar part, the exposed interface has regions associated with the other puzzle 
bars. The model cannot be partitioned unless this exact region attribute pattern is reproduced; this pattern cannot be reproduced by our system. We expect such pathological cases to be rare.

\revised{Finally, a number of choices in our region extractability computation (limiting intersection testing to vertices, fixing the number of  sampled extraction directions) are guided by our desire to achieve an acceptable performance versus robustness tradeoff. We have never seen failures emanating from 
either choice, but they are theoretically possible.  Increasing the number of directions sampled or using robust predicates for intersection tests would make the method more fail-proof, but will likely significantly slow it down. Robustness is also limited by the precision 
allowable in the interior tetrahedral discretization, as is the case for all discrete frameworks.}

\begin{acks}
The authors wish to thank the reviewers for their insightful suggestions. We are also deeply grateful to: Enrique Rosales, for help with model creation; Jinfan Yang, Linda Lin Lu, Riccardo Scateni, Gianmarco Cherchi, Stefano Nuvoli and Alessandro Tola, for help with 3D printing and milling; Michela Spagnuolo, for early discussions on this project; the authors of~\cite{yao2017interactive}, for running their algorithm on our input data. The UBC authors were supported by NSERC. The CNR IMATI authors were supported by the EU Horizon 2020 program, under grant agreement No.680448 (CAxMan).
\end{acks}

\bibliographystyle{ACM-Reference-Format}
\bibliography{00_main}


\begin{appendices}

\section{Part Extractability}
\label{app:proof}

Without loss of generality, we assume that all the parts $\Obj_j$ with $j<i$ have already been extracted.

\begin{prop} A part $\Obj_i$ is extractable along a direction $d_i$ if it is both \emph{interface extractable} and \emph{\old{surface}\revised{region} extractable}.
\label{prop:extractable}
\end{prop}

\paragraph{Proof: } Any point $p \in \Obj_i$ must be in one of the following conditions: (1) on the surface of $\Obj$ but not on the interface with other parts; (2) on the interface with other parts; (3) neither of the previous two (i.e. internal). Considering a ray shot from $p$ along $d_i$, we have the following cases:
\\
\ if (1), the ray cannot intersect the boundary of any $\Obj_j$ with $j>i$ because $\Obj_i$ is \emph{\old{surface}\revised{region} extractable}. Hence in order to intersect a part $\Obj_j$ with $j > i$, the ray must necessarily exit $\Obj_i$ to enter $\Obj_j$ at a common interface point which is not on the boundary of $\Obj$. However, since the ray \emph{exits} $\Obj_i$, this would require that this common point on the boundary of $\Obj_i$ has a normal which is coherently oriented with $d_i$, which is excluded because we assume \emph{interface extractability}. Hence, no ray originated on a boundary point can intersect a part $\Obj_j$ with $j > i$;
\\
 if (2), the ray necessarily enters $\Obj_i$ because we assume \emph{interface extractability}. This means that, before possibly intersecting any other $\Obj_j$ with $j > i$, it must exit $\Obj_i$ from a point on the boundary of $\Obj$, and this leads us back to the previous case;
\\
if (3), before possibly intersecting any other $\Obj_j$ with $j > i$, the ray must exit $\Obj_i$, reducing to one of the previous two cases.

Thus, any possible ray reduces to case (1); this in turn implies that no ray from $\Obj_i$ along $d_i$ intersects other parts $\Obj_j$ with $j>i$. \qedsymbol

\end{appendices}


\end{document}